\RequirePackage{fix-cm}
\documentclass[smallextended]{svjour3}       % onecolumn (second format)
\smartqed  % flush right qed marks, e.g. at end of proof
\usepackage{graphicx}

\usepackage{amssymb}

\usepackage{longtable}

\usepackage[numbers]{natbib}

\usepackage{booktabs}
\usepackage{multirow}
\usepackage{listings}
\usepackage{tikz}
\usetikzlibrary[positioning]
\usetikzlibrary{patterns}

\newcommand{\boxwidth}{4.5 cm}
\newcommand{\boxtextwidth}{4.5 cm}
\newcommand{\classificationheight}{1 cm}

\usepackage{array}
\usepackage{pxfonts}

\usepackage{csquotes}
\usepackage{hyperref}

\usepackage{geometry}
\usepackage{pdflscape}

\usepackage{graphicx}
\usepackage{epstopdf} % for eps graphics
\usepackage{subcaption}
\usepackage[ruled]{algorithm2e}

\SetAlFnt{\small}
\SetAlCapFnt{\small}
\SetAlCapNameFnt{\small}
\SetAlCapHSkip{0pt}
\IncMargin{-\parindent}

\begin{document}

\title{Using Meta-heuristics and Machine Learning for Software Optimization of Parallel Computing Systems: A Systematic Literature Review}

\titlerunning{Using Meta-heuristics and Machine Learning for Software Optimization of Parallel Computing Systems}        % if too long for running head

\author{Suejb Memeti        \and
        Sabri Pllana		\and
        Al\'{e}cio Binotto	\and
        Joanna Ko\l{}odziej	\and
        Ivona Brandic
}

\authorrunning{S. Memeti, S. Pllana, A. Binotto, J. Ko\l{}odziej, and I. Brandic} % if too long for running head

\institute{		
				S. Memeti \and S. Pllana \at
			  	Linnaeus University, Department of Computer Science, 351 95 V\"{a}xj\"{o}, Sweden \\
              	\email{suejb.memeti@lnu.se} \\           %  \\
              	\email{sabri.pllana@lnu.se}
			\and
				A. Binotto \at
				IBM Research, Brazil\\
				\email{abinotto@br.ibm.com}
			\and
				J. Ko\l{}odziej \at
				Cracow University of Technology, 31 155 Cracow, Poland\\
				\email{jokolodziej@pk.edu.pl}
			\and	
				I. Brandic \at
				Electronic Commerce Group, Institute of Software Technology and Interactive Systems, Vienna University of Technology, A-1040 Vienna, Austria
				\email{ivona.brandic@tuwien.ac.at}
}

\date{PREPRINT}

\maketitle

\begin{abstract}
While modern parallel computing systems offer high performance, utilizing these powerful computing resources to the highest possible extent demands advanced knowledge of various hardware architectures and parallel programming models. Furthermore, optimized software execution on parallel computing systems demands consideration of many parameters at compile-time and run-time. Determining the optimal set of parameters in a given execution context is a complex task, and therefore to address this issue researchers have proposed different approaches that use heuristic search or machine learning. In this paper, we undertake a systematic literature review to aggregate, analyze and classify the existing software optimization methods for parallel computing systems. We review approaches that use machine learning or meta-heuristics for software optimization at compile-time and run-time. Additionally, we discuss challenges and future research directions. The results of this study may help to better understand the state-of-the-art techniques that use machine learning and meta-heuristics to deal with the complexity of software optimization for parallel computing systems. Furthermore, it may aid in understanding the limitations of existing approaches and identification of areas for improvement.


\end{abstract}

% ================================================================================================= %
\section{Introduction}
\label{sec:introduction}

Traditionally, parallel computing \citep{Padua:2011:EPC} systems have been used for scientific and technical computing. Usually scientific and engineering computational problems are complex and resource intensive. To efficiently solve these problems, utilization of parallel computing systems that may comprise multiple processing units is needed. The emergence of multi-core and many-core processors in the last decade led to the pervasiveness of parallel computing systems from embedded systems, personal computers, to data centers and supercomputers. While in the past parallel computing was a focus of only a small group of scientists and engineers at supercomputing centers, nowadays programmers of virtually all systems are exposed to parallel processors that comprise multiple or many cores \citep{jeffers2015high}.

The modern parallel computing systems offer high performance capabilities. In recent years, the computational capabilities of supercomputing centers have been increasing very fast. For example, the average performance of the top 10 supercomputers in 2010 was 0.84 PFlops/s, in 2014 the average performance climbed to 11.16 PFlops/s, and in 2016 the average performance capability is 20.63 PFlops/s \citep{top500}. With such exciting performance gain, a serious issue of the power consumption of these supercomputing centers arises. For example, according to the TOP 500 list \citep{top500}, in the years 2010 to 2016, the average power consumption of the top 10 supercomputers has increased from 2.98MW to 8.88MW, that is about 198\% increase.

Utilizing these resources to gain the highest extent of performance while keeping low level of energy consumption demands significant knowledge of vastly different parallel computing architectures and programming models. Improving the resource utilization of parallel computing systems (including heterogeneous systems that comprise multiple non-identical processing elements) is important, yet difficult to achieve \cite{jin2016surveyenergy}. For example, for data-intensive applications the limited bandwidth of the PCIe interconnection forces developers to use the resources on the host only, which leads to the underutilization of the system. Similarly, in compute-intensive applications, while utilizing the accelerating device, the host CPUs remain idle, which leads to waste of energy and performance. Approaches that intelligently manage the resources of host CPUs and accelerating devices to address such inefficiencies seem promising \citep{mittal2015survey}.

To achieve higher performance, scalability and energy efficiency, engineers often combine Central Processing Units (CPUs), Graphical Processing Units (GPUs), or Field Programmable Gate Arrays (FPGAs). In such environments, system developers need to consider multiple execution contexts with different programming abstractions and run-time systems. There is a consensus that software development for parallel computing systems, especially heterogeneous systems, is significantly more complex than for traditional sequential systems. In addition to the programmability challenges, performance portability of programs to various platforms is essential and challenging for productive software development, due to the differences in architectural level of multi-core and many-core processors \citep{benkner2011peppher}.

Software development and optimal execution on parallel computing systems expose programmers and tools to a large number of parameters \citep{sandrieser12} at software compile-time and at run-time. Examples of properties for a GPU-accelerated system include: CPU count, GPU count, CPU cores, CPU core architecture, CPU core speed, memory hierarchy levels, GPU architecture, GPU device memory, GPU SM count, CPU cache, CPU cache line, memory affinity, run-time system, etc. Finding the optimal set of parameters for a specific context is a non-trivial task, and therefore many methods for software optimization that use meta-heuristics and machine learning have been proposed. A systematic literature review may help to aggregate, analyze, and classify the proposed approaches and derive the major lessons learned.

In this paper, we conduct a systematic literature review of approaches for software optimization of parallel computing systems. We focus on approaches that use machine learning or meta-heuristics that have been published since the year 2000. We classify the selected review papers based on the software life-cycle activities (compile-time or run-time), target computing systems, optimization methods, and period of publication. Furthermore, we discuss existing challenges and future research directions. The aims of this systematic literature review are to:
\begin{itemize}
	\item systematically study the state-of-the-art software optimization methods for parallel computing systems that use machine learning or meta-heuristics;
	\item classify the existing studies based on the software life-cycle activities (compile-time, and run-time), target computing systems, optimization methods, and period of publication;
	\item discuss existing challenges and future research directions.
\end{itemize}

\begin{figure}[t]
	\centering
	\includegraphics[width=\linewidth]{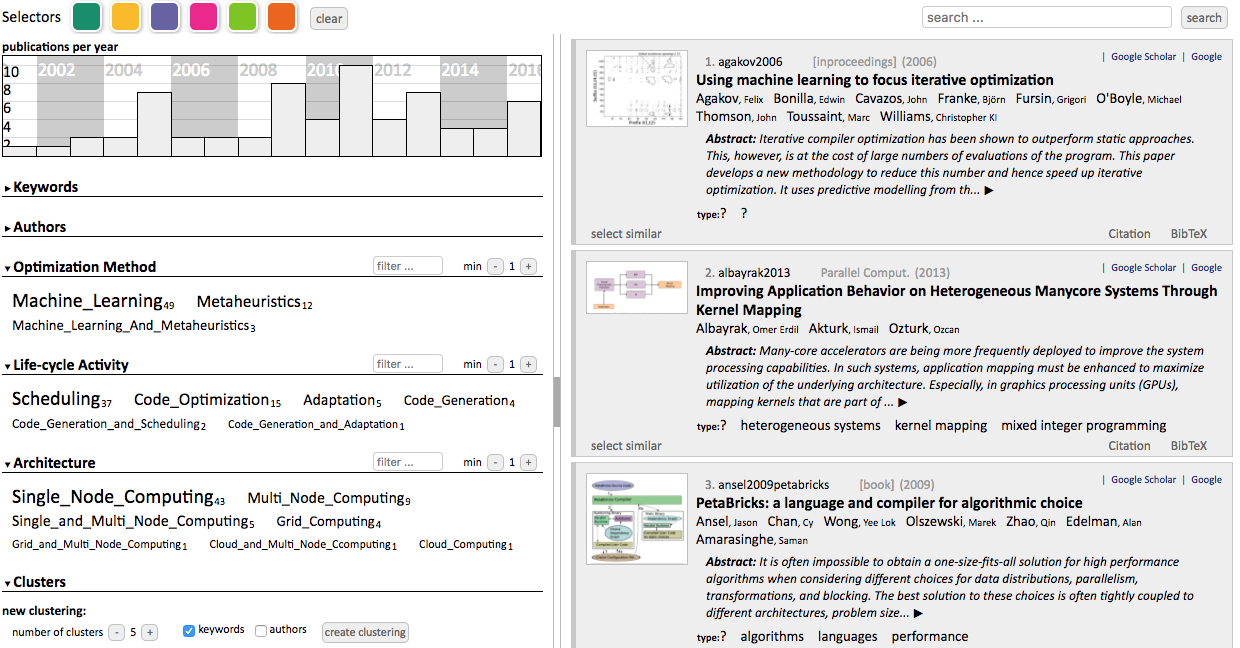}
	\caption{Our interactive browser of the results of literature review. Results can be filtered by software optimization method, software life-cycle activity, parallel system architecture, keyword, and author name. Results are visualized in the form of time-line that indicates the number of publications per year. The right-hand compartment lists the publications that match the search criteria. The browser is available on-line at \url{www.smemeti.com/slr/}.}
	\label{fig:survis}
\end{figure}

Figure \ref{fig:survis} depicts our solution for browsing the results of literature review that we have developed using SurVis \cite{Beck2016survis} literature visualization tool. The browser is available on-line at \url{www.smemeti.com/slr/} and enables to filter the review results based on the optimization methods, software life-cycle activity, parallel computing architecture, keywords, and authors. A time-line  visualizes the number of publications per year. Publications that match filtering criteria are listed on the right-hand side; the browser displays for each publication the title, authors, abstract, optimization method, life-cycle activity, target system architecture, keywords, and a representative figure. The on-line literature browser is easy to extend with future publications that fit the scope of this review.

The rest of the paper is organized as follows. In section \ref{sec:research-method} we describe the research methodology. In section \ref{sec:taxonomy-terminology}, we give an overview of the parallel computing systems, software optimization techniques, and the software optimization at different life-cycle activities. For each of the software life-cycle activities, including \emph{Compile-Time} activities (Section \ref{sec:compile-time}), and \emph{Run-Time} activities (Section \ref{sec:run-time}), we discuss the characteristics of state-of-the-art research, and discuss limitations and future research directions. Finally, in Section \ref{sec:conclusion} we conclude our paper.

% ================================================================================================= %

\section{Research methodology}
\label{sec:research-method}

\begin{figure}[ht]
	\centering
	\includegraphics[width=.8\linewidth]{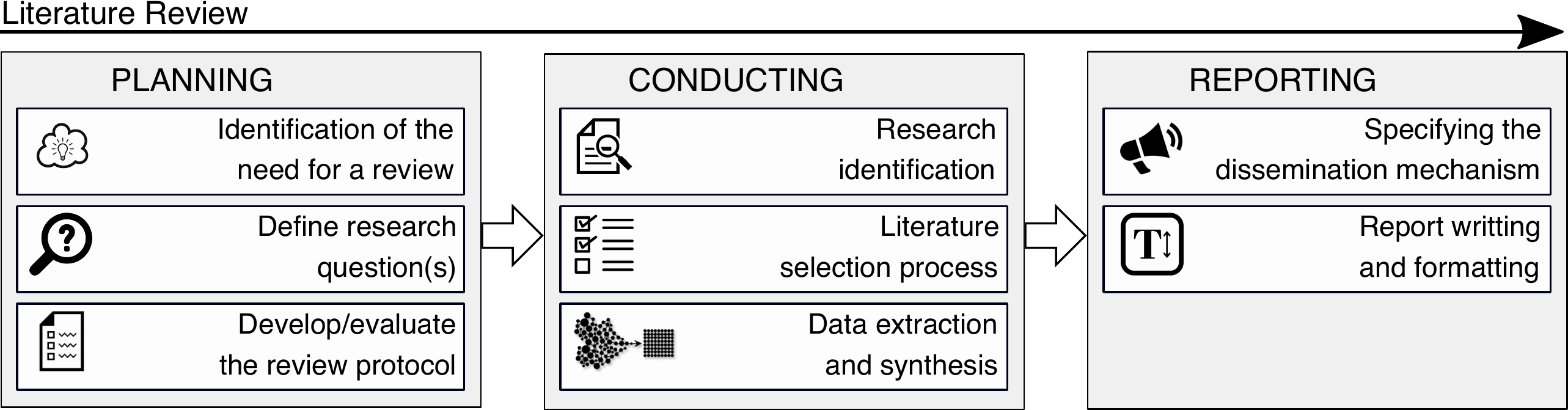}
	\caption{Research methodology}
	\label{fig:research_methodology}
\end{figure}

We perform a literature review based on guidelines by \citet{Kitchenham07GPSL}. In summary, these guidelines include three stages: Planning, Conducting and Reporting (see Fig. \ref{fig:research_methodology}).

During the planning stage the following activities are performed: (1) identifying the need for a literature review, (2) defining the research questions of the literature review, and (3) developing/evaluating the protocol for performing the literature review.
The activities associated with conducting the literature review include: (1) identifying the research, (2) literature selection, (3) data extraction and synthesis.
The reporting stage includes writing the results of the review and formatting the document. In what follows, we describe in more details the research method and the major activities performed during this study.

% ------------------------------------------------------------------------------------------------- %
\subsection{Research questions}
\label{sec:research-questions}

We have defined the following research questions:

\noindent
\fbox{\parbox{\textwidth}{
		\begin{itemize}
			\item \textbf{\emph{RQ1}}: Which software optimization goals for parallel computing systems are achieved using meta-heuristics and machine learning?
			\item \textbf{\emph{RQ2}}: Which are the common algorithms used to achieve such software optimization goals for parallel computing systems?
			\item \textbf{\emph{RQ3}}: Which features are considered during software optimization of parallel computing systems?
		\end{itemize}
}}

% ------------------------------------------------------------------------------------------------- %
\subsection{Search and Selection of Literature}

\begin{figure}[t]
	\centering
	\includegraphics[width=\linewidth]{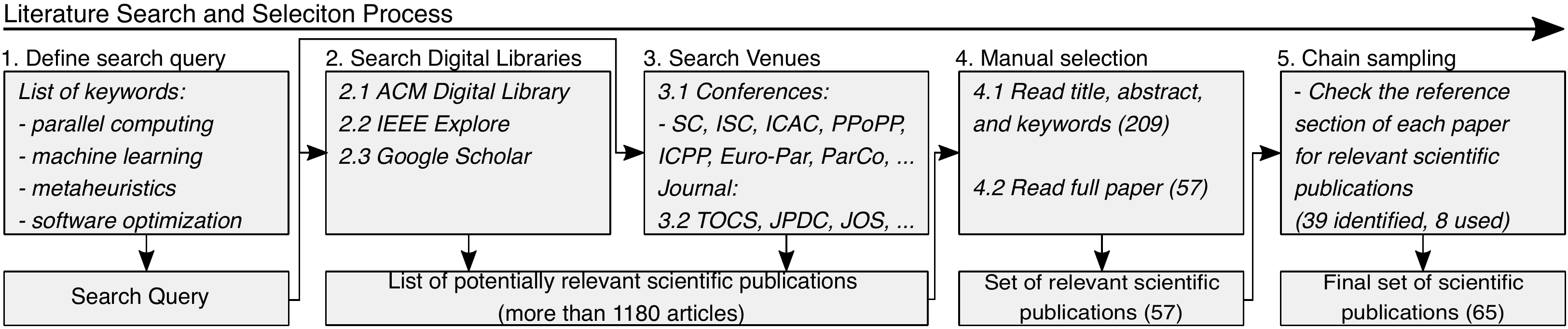}
	\caption{The process of searching and selecting the relevant literature}
	\label{fig:literature_search_and_select}
\end{figure}

The literature search and selection process are depicted in Fig. \ref{fig:literature_search_and_select}. Based on the objectives of the study, we have selected an initial set of keywords (see activity 1) that is used to search for articles, such as: \emph{parallel computing, machine learning, meta-heuristics} and \emph{software optimization}. To improve the result of the search process, we consider synonyms for the keywords during the search. The search query is executed on digital electronic databases (such as, ACM Digital Library, IEEEXplore, and Google Scholar), conference venues (such as, SC, ISC, ICAC, PPoPP, ICDCS, CGO, ICPP, Euro-Par, and ParCo), and scientific journals (such as, TOCS, JPDC, JOS). The outcome of the search process is a list of potentially relevant scientific publications. Manual selection of these publications by reading the title, abstract, and keywords (activity 4.1) first, then the full paper (activity 4.2) is performed, which results in a filtered list of relevant scientific publications. Furthermore, a recursive procedure of searching for related articles is performed using the corresponding related articles section of each digital library (for example, the ACM Digital Library related papers function powered by IBM Watson, or the \emph{Related articles} function of Google Scholar).

The initial automatic search on the ACM digital library (see Figure \ref{fig:literature_search_and_select}, activity 2.1) returned a list of total 25970 entries (articles). We sorted the entries by relevance, such that the most relevant articles will show up first. As expected, the most relevant articles were found in the first part of the list, and after hundreds of articles, the suggested entries were not relevant to our study. Therefore, we decided to consider only the first 1000 articles. Out of these articles, only 130 were selected for further study based on reading the title and abstract (activity 4.1), and after reading the full article (activity 4.2), 22 were selected as relevant articles. The IEEEXplore returned 40 potentially relevant articles (activity 2.2), 20 of them were selected for further study based on reading the title and abstract (activity 4.1), and 16 were selected as relevant after reading the full paper (activity 4.2). The Google Scholar returned 140 potentially relevant articles (activity 2.3), 31 of them were selected after reading the title and abstract (activity 4.1), and 11 were selected as relevant after reading the full paper (activity 4.2). Searching the conference venues (activity 3.1) and scientific journals (activity 3.2), we selected 28 articles based on reading the title and abstract (activity 4.1), and 8 of them were selected as relevant after reading the full paper (activity 4.2). So, out of more than 1180 articles returned from various sources (activity 2 and 3), 209 were selected manually based on reading the title and abstract (activity 4.1), out of which, after reading the full content (activity 4.2), 57 were selected as relevant to the scope of this paper.

Additionally, the chain sampling technique (also known as snowball sampling, see Figure \ref{fig:literature_search_and_select}, activity 5) is used to search for related articles. 39 articles were identified using this technique by reading the title and abstract (activity 4.1), and 8 of them were selected as relevant after reading the full paper (activity 4.2). Chain sampling is a recursive technique that considers existing articles, usually found in the references section of the research publication under study \cite{biernacki1981snowball}. In total, 65 publications are considered in this review.

% ------------------------------------------------------------------------------------------------- %
\subsection{The Focus and Scope of the Literature Review (Selection Process)}
\label{sec:focus}

The scope of this literature review includes:

\begin{itemize}
	\item publications that investigate the use of machine learning or meta-heuristics for software optimization of parallel computing systems;
	\item publications that contribute to compile-time activities (code optimization and code generation), and run-time activities (scheduling and adaptation) of software life-cycle;
	\item research published since the year 2000, because in literature, the year 2000 is considered as the starting point of the multi-core era. IBM Power 4 \cite{diefendorff1999power4}, the first industry dual-core processor, is introduced in 2001 \cite{geer2005}.
\end{itemize}

While other optimization methods (such as, linear programming, dynamic programming, control theory), and other software optimization activities (such as, design-time software optimization) may be of interest, they are left out of scope to keep the systematic review focused.

% ------------------------------------------------------------------------------------------------- %
\subsection{Data Extraction}
\label{sec:data_extraction}

In accordance with the classification strategy (described in Section \ref{sec:optimization-activities}) and the defined research questions (described in Section \ref{sec:research-questions}), for each of the selected primary studies we have collected information that we consider important to be recorded in order to perform the literature review.

Table \ref{tab:data-extraction} shows an excerpt of the data items (used for quantitative and qualitative analysis) collected for each of the selected studies. Data items 1-3 are used for the quantitative analysis related to \emph{RQ1}. Data item 4 is used to answer \emph{RQ2}. Data collected for item 5 is used to answer \emph{RQ3}, whereas data collected for item 6 is used to answer \emph{RQ4}. Data item 7 is used to classify the selected scientific publications based on the software life-cycle activities (see Table \ref{tab:classification-software-lifecycle}), whereas data item 8 is used for the classification based on the target architecture (see Fig. \ref{fig:classification-architecture}).

\begin{table}
	\centering
	\scriptsize
	%	\footnotesize
	\caption{An excerpt of data items collected for each of the selected publications.}
	\label{tab:data-extraction}
	\begin{tabular}{l p{3.2cm} p{10cm}}
		\toprule
		&   Data item							&	Description \\ \midrule
		1	&   Date							&	Date of the data extraction\\
		2	&	Bibliographic reference			&	Author, Year, Title, Research Center, Venue \\
		3	&	Type of article					&	Journal article, conference paper, workshop paper, book section \\
		4	&	Problem, objectives, solution	&	What is the problem; what are the objectives of the study; how the proposed solution works? \\
		5	&	Optimization Technique			&	Which Machine Learning or Meta-heuristic algorithm is used? \\
		6	&	Considered features				&	The list of considered features used for optimization \\
		7	&	Life-cycle Activity				&	Code Optimization, Code Generation, Scheduling, Adaptation?	 \\
		8	&	Target architecture				&	Single/Multi-node system, Grid Computing, Cloud Computing\\
		9	&	Findings and conclusions 		&	What are the findings and conclusions? \\
		10	&	Relevance						&	Relevance of the study in relation to the topic under consideration \\
		%&	...								&	... \\
		\bottomrule
	\end{tabular}
\end{table}

% ================================================================================================= %
\section{Taxonomy and Terminology}
\label{sec:taxonomy-terminology}

In this section, we provide an overview of the parallel computing systems and software optimization approaches with focus on machine learning and meta-heuristics. Thereafter, we present our approach for classifying the state-of-the-art optimization techniques for parallel computing.

% ------------------------------------------------------------------------------------------------- %
\subsection{Parallel Computing Systems}
\label{sec:parallel-systems}

A parallel computing system comprises a set of interconnected processing elements and memory modules. Based on the system architecture, generally parallel computers can be categorized into shared and distributed memory. \textit{Shared memory} parallel computing systems communicate through a global shared memory, whereas in \textit{distributed memory} systems every processing element has its own local memory and the communication is performed through message passing. While shared memory systems have shown limited scalability, distributed memory systems have demonstrated to be highly scalable. Most of the current parallel computing systems use shared memory within a node, and distributed memory between nodes \citep{barney2010introduction}.

According to Top500 \citep{top500} in the 90s the commonly used parallel computing systems were \textit{symmetric multi-processing} (SMP) systems and \textit{massive parallel processing} (MPP) systems. SMPs are shared memory systems where two or more identical processing units share other system resources (main memory, I/O devices) and are controlled by a single operating system. MPPs are distributed memory systems where a larger number of processing units (or separate computers) are housed in the same place. The disparate processing units share no system resources, they have their own operating system, and communicate through high-speed network. The main computing models within the distributed parallel computing systems include \textit{cluster} \cite{Sterling95beowulf,dongarra2005high}, \textit{grid} \cite{smanchat2013,buyya2009,foster2003,sadashiv2011}, and \textit{cloud} computing \cite{sadashiv2011,foster2008,malawski2015}.

Nowadays, the mainstream platforms for parallel computing, at their node level consist of multi-core and many-core processors. \textit{Multi-core} processors may have multiple cores (two, four, eight, twelve, sixteen...) and are expected to have even more cores in the future. \textit{Many-core} systems consist of larger number of cores. The individual cores of the many-core systems are specialized to efficiently perform operations such as, SIMD, SIMT, speculations, and out-of-order execution. These cores are more energy efficient because they usually run at lower frequency.

Systems that comprise multiple identical cores or processors are known as \textit{homogeneous systems}, whereas \textit{heterogeneous systems} comprise non-identical cores or processors. As of November 2017, the TOP500 list \citep{top500} contains several supercomputers that comprise multiple heterogeneous nodes. For example, a node of Tianhe-2 (2nd most powerful supercomputer) comprises Intel Ivy-Bridge multi-core CPUs and Intel Xeon Phi many-core accelerators; Piz Daint (3rd) consists of Intel Xeon E5 multi-core CPUs and NVIDIA Tesla P100 many-core GPUs \citep{memeti2017benchmarking,viebke2015}.

Programming parallel computing systems, especially heterogeneous ones, is significantly more complex than programming sequential processors \citep{PllanaBMNX08}. Programmers are exposed to various parallel programming languages (often implemented as extensions of general-purpose programming languages such as C and C++), including, OpenMP \citep{openmp2013}, MPI \citep{gropp1999using}, OpenCL \citep{opencl2010stone}, NVIDIA CUDA \citep{cuda}, OpenACC \citep{Wienke:2012} or Intel TBB \citep{tbb:2011}. Additionally, the programmer is exposed to different architectures with different characteristics (such as the number of CPU/GPU devices, the number of cores, core speed, run-time system, memory and memory levels, cache size). Finding the optimal system configuration that results in the highest performance is challenging. In addition to the programmability challenge, heterogeneous parallel computing systems bring the portability challenge, which means that programs developed for a processor architecture (for instance, Intel Xeon Phi) may not function on another processor architecture (such as, GPU). Manual software porting and performance tuning for various architectures may be prohibitive.

Existing approaches, discussed in this study, propose several solutions that use machine learning or meta-heuristics during compile-time and run-time to alleviate the programmability and performance portability challenges of parallel computing systems.

% ------------------------------------------------------------------------------------------------- %
\subsection{Software Optimization Approaches}
\label{sec:software-optimization}

In computer science selecting the best solution considering different criteria from a set of various available alternatives is a frequent need. Based on what type of values the model variables can take, the optimization problems can be broadly classified in continuous and discrete. Continuous optimization problems are concerned with the case where the model variables can take any value permitted by some given constraints. Continuous optimization problems are easier to solve. Given a point $x$, using continuous optimization techniques one can infer information about neighboring points of $x$ \citep{gould2006introduction}.

In contrast, in discrete optimization (also known as combinatorial optimization) methods the model variables belong to a discrete set (typically subset of integers) of values. Discrete optimization deals with problems where we have to choose an optimal solution from a finite number of possibilities. Discrete optimization problems are usually hard to solve and only enumeration of all possible solutions is guaranteed to give the correct result. However, enumerating across all available solutions in a large search space is prohibitively demanding.

Heuristic-guided approaches are designed to solve optimization problems more quickly by finding approximate solutions when other methods are too slow or fail to find any exact solution. These approaches select near-optimal solutions within a time frame (that is, they trade-off optimality for speed). While heuristics are designed to solve a particular problem (problem-dependent), meta-heuristics can be applied to a broad range of problems. They can be thought as higher-level heuristics that are designed to determine a near-optimal solution to an optimization problem, with limited computation capacity and knowledge about the problem.

In what follows, we first describe the meta-heuristics and list commonly used algorithms, and thereafter, we describe machine learning in the context of software optimization.

\begin{figure}[ht]
	\centering
	\includegraphics[width=.7\linewidth]{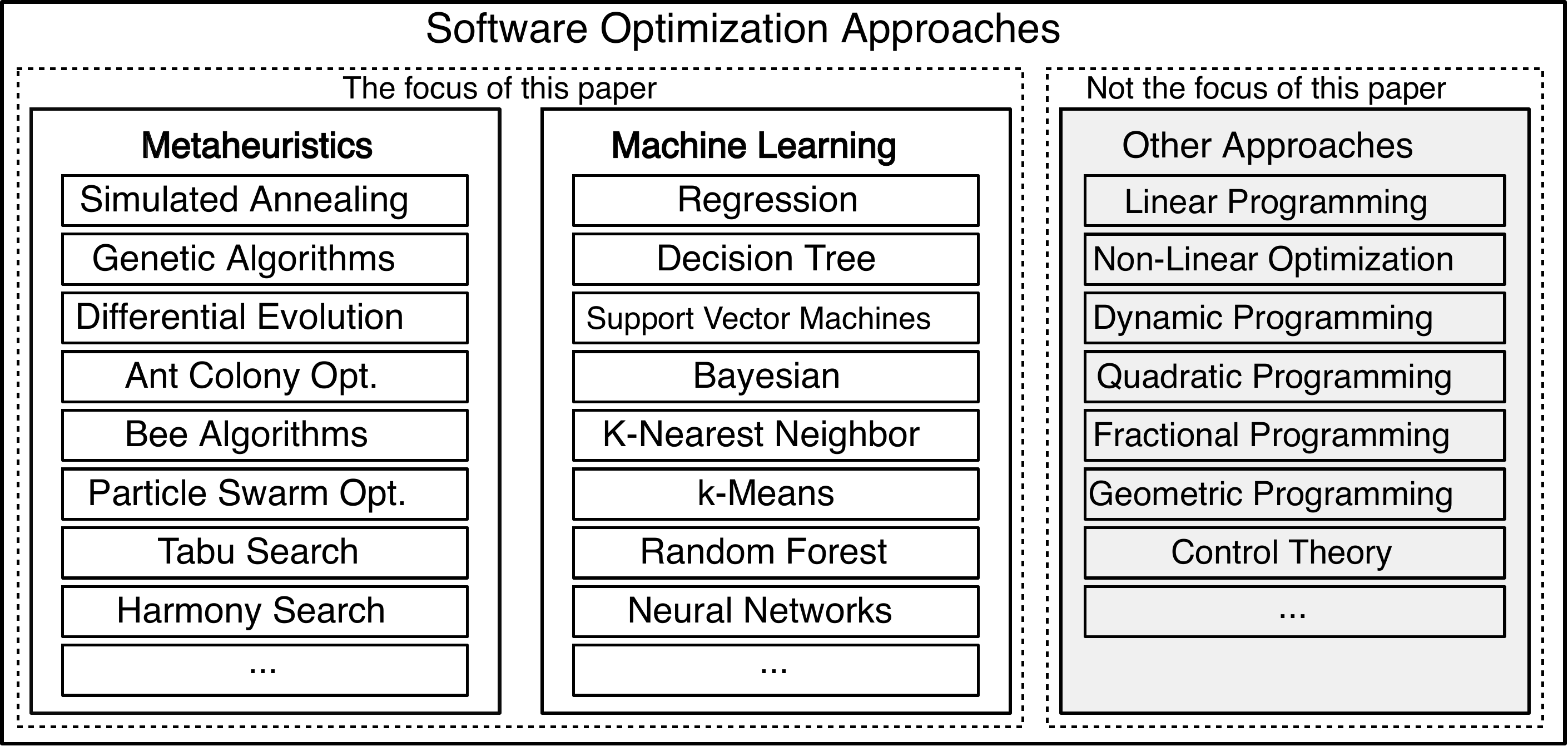}
	\caption{Classification of the software optimization approaches. While there exist many different optimization approaches, in this study we focus on meta-heuristics and machine learning.}
	\label{fig:optimization-approaches}
\end{figure}

% ------------------------------------------------------------------------------------------------- %
\subsubsection{Meta-heuristics}
\label{sec:meta-heuristics}

Meta-heuristics are high-level algorithms that are capable to determine a sufficiently satisfactory (near-optimal) solution to an optimization problem with limited domain knowledge and computation capacity. As meta-heuristics are problem-independent they can be used for a variety of problems. Meta-heuristics algorithms are often used for the management and efficient use of resources to increase productivity \citep{NumRecipes2007,wolsey2014integer}. In cases where the search space is large, exhaustive search, iterative methods, or simple heuristics are impractical, whereas meta-heuristics can often find good solutions with less computational effort. Meta-heuristics have shown to provide efficient solution to different problems, such as the minimum spanning tree (MST), traveling salesman problem (TSP), shortest path trees, and matching problems. Selecting the most suitable heuristic for a specific problem is important to reach a near-optimal solution more quickly. However, this process requires consideration of various factors, such as the domain type, search space, computational time, and solution quality \citep{memeti2016bd,braun2001comparison}.

In the context of software optimization, the commonly used meta-heuristics include Genetic Algorithms, Simulated Annealing, Ant Colony Optimization, Local Search, Tabu Search, and Particle Swarm Optimization (see Figure~\ref{fig:optimization-approaches}).

% ------------------------------------------------------------------------------------------------- %
\subsubsection{Machine Learning}
\label{sec:ml}

Machine Learning is a technique that allows computing systems to learn (that is, improve) from the experience (available data). \citet{Mitchell:1997} defines Machine Learning as follows, \enquote{A computer program is said to learn from experience \emph{E} with respect to some class of tasks \emph{T} and performance measure \emph{P}, if its performance at tasks in \emph{T}, as measured by \emph{P}, improves with experience \emph{E}}.

Machine learning programs operate by building a prediction model from a set of training data, which later on is used to make data-driven predictions, rather than following hard-coded static instructions. Some of the most popular machine learning algorithms (depicted in Fig. \ref{fig:optimization-approaches}) include regression, decision tree, support vector machines, Bayesian inference, random forest, and artificial neural networks.

An important process while training a model is the \emph{feature selection}, because the efficiency of models depends on the selected variables. It is critical to choose features that have significant impact on the prediction model. There are different feature selection techniques that can find features that contain the most useful information to distinguish between classes, for example mutual information score (MIS) \citep{duda1973}, greedy feature selection \citep{stephenson2005}, or information gain ratio \citep{guyon2003}.

Depending on the way the prediction model is trained, machine learning may be \textit{supervised} or \textit{unsupervised}. In supervised machine learning the prediction model learns from examples that are labeled, which means that the input and the output are known in the training data set. Supervised learning uses classification techniques to predict discrete responses (such as, determining whether an e-mail is genuine or spam, determining whether a tumor is malign or benign), and regression techniques to predict continuous responses (such as, changes in temperature, fluctuations in power demand). The most popular supervised learning algorithms for classification problems include Support Vector Machines, Naive Bayes, Nearest Neighbor, and Discriminant Analysis, whereas for regression problems algorithms such as Linear Regression, Decision Trees, and Neural Networks are used. Selecting the best algorithm depends on the size and type of input data set, the desired output (insight), and how those insights will be used.

The unsupervised machine learning models have no or very little knowledge of how the results should look like. Basically, correct results (that is labeled training data sets) are not used for model training, but the model aims at finding hidden patterns in data based on statistical properties (for instance, intra-cluster variance) of the training data sets. Unsupervised learning can be used for solving data clustering problems in various domains, for example, sequence analysis, market research, object recognition, social network analysis, and astronomical data analysis. Some commonly used algorithms for data clustering include K-Means, Hierarchical Clustering, Neural Networks, Hidden Markov Model, and Density-based Clustering.

% ------------------------------------------------------------------------------------------------- %
\subsection{Software Optimization at Different Software Life-cycle Activities}
\label{sec:optimization-activities}

\begin{figure}[ht]
	\centering
	\includegraphics[width=.7\linewidth]{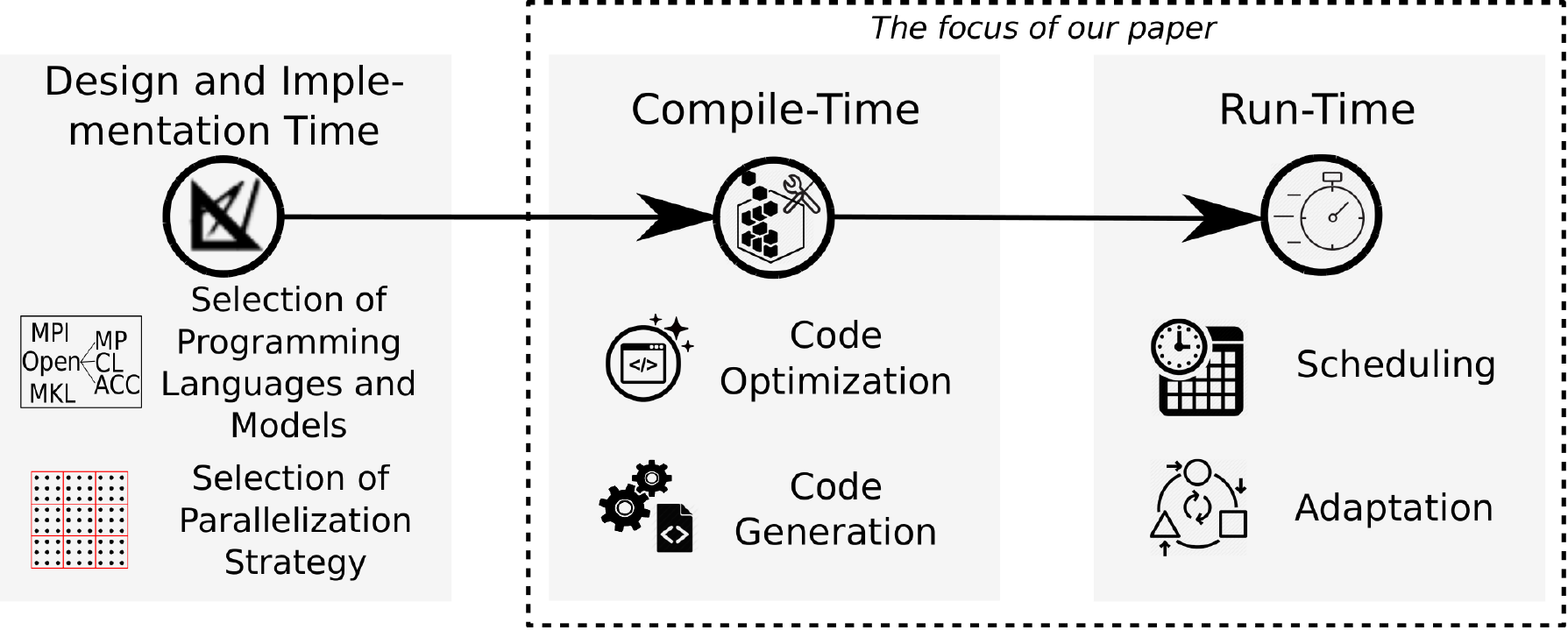}
	\caption{Software life-cycle activities. At design and implementation the selection of the programming languages, models, and the parallelization strategy occurs. We focus on software optimization that occurs during software compile-time (that includes code optimization and generation), and during run-time (that includes scheduling and adaptation).}
	\label{fig:so-time}
\end{figure}

Software optimization can happen during different activities of the software life-cycle. We categorize the software optimization activities by the time of their occurrence: \emph{Design and Implementation-time}, \emph{Compile-time}, \emph{Run-time} (Fig. \ref{fig:so-time}).

During the \emph{design and implementation} activity, decisions such as selection of the programming language/model and selection of the parallelization strategy are considered.

The \emph{compile-time} activities include decisions of selecting the optimal compiler optimization flags and source code transformations (such as loop unrolling, loop nest optimization, pipelining, and instruction scheduling) such that the executable program is optimized to achieve certain goals (performance or energy) on a given context.

The \emph{run-time} activities include decisions of selecting the optimal data and task scheduling on parallel computing systems, as well as taking decisions (such as switching to another algorithm or changing the clock frequency) that help the system to adapt itself during the program execution and improve the overall performance and energy efficiency.

While software design and implementation activities are performed by the programmer, software activities at compile-time and run-time are completed by tools (such as compilers and run-time systems). Therefore, in this paper we focus on tool-supported software optimization approaches that use approximate techniques (machine learning and meta-heuristics) at compile-time and run-time.

For each of the software optimization life-cycle activities, including \emph{Compile-Time} (section \ref{sec:compile-time}) and \emph{Run-Time} (section \ref{sec:run-time}), we will describe the context for software optimization goals, discuss the state-of-the-art research, and discuss limitations and future research directions.

% ------------------------------------------------------------------------------------------------- %
\subsection{Classification based on architecture, software optimization approach, and life-cycle activity}
\label{sec:classification}

In this section we classify the considered scientific publications based on the architecture, software optimization approach, and life cycle activities.

To provide an overview of the current state of the art, we have grouped the scientific publications that use machine learning and meta-heuristics for software optimization of parallel computing systems in the following time periods: 2000-2005, 2006-2011, and 2012-2017. Each of the periods, correspond to the type of the processors that were used the most in the TOP list during that time. For example, even though the first multi-core processor was introduced in 2001 \cite{geer2005}, most of the super computers in TOP500 list during years 2000-2005 comprised multiple single-core processors \cite{top500}. Further filtering and classification of the considered scientific publications, and visualization of the results in the form of a time-line can be performed using our on-line interactive tool (see Fig. \ref{fig:survis}).

\begin{figure}[ht]
	\centering
	\scriptsize
	
	\begin{tikzpicture}[x=\boxwidth, y=-1cm, node distance=0 cm,outer sep = 0pt]
	% Style for x-axis
	\tikzstyle{year}=[draw, rectangle,  minimum height=1cm, minimum width=\boxwidth, color=white, text=black, anchor=south west]
	\tikzstyle{no-border}=[draw, rectangle,  minimum height=1cm, minimum width=\boxwidth, draw=none, text=black, anchor=south west]
	% Style for y-axis
	\tikzstyle{classification}=[draw, rectangle, minimum height=\classificationheight, minimum width=1 cm, color=white, text=black, anchor=north east]

	% Styles for events
	% Duration of sequences
	\tikzstyle{classifications}=[rectangle,draw, minimum width=\boxwidth, anchor=north west,text centered,text width=\boxtextwidth]
	\tikzstyle{1hour}=[classifications,minimum height=\classificationheight]
	
	%Style for type of sequence
	\tikzstyle{ml-cell}=[1hour,fill=white!20]
	\tikzstyle{mh-cell}=[1hour,fill=white!20]
	
	% Positioning labels for days and hours
	\node[year] (2000-2005) at (0,5) {\footnotesize 2000-2005};
	\node[year] (2006-2011) [right = of 2000-2005] {\footnotesize 2006-2011};
	\node[year] (2012-2017) [right = of 2006-2011] {\footnotesize 2012-2017};
	\node[classification] (CO) at (0,0) {\scriptsize \rotatebox{90}{\parbox{.8cm}{\centering Cloud \\Comp.}}};
	\node[classification] (CG) [below = of CO] {\scriptsize \rotatebox{90}{\parbox{.8cm}{\centering Grid \\ Comp.}}};
	\node[classification] (SC) [below = of CG] {\scriptsize \rotatebox{90}{\parbox{.8cm}{\centering Single-Node}}};
	\node[classification] (AD) [below = of SC] {\scriptsize \rotatebox{90}{\parbox{.8cm}{\centering Multi-Node}}};
	
	%		\shade[left color=white,right color=white] (0,0) rectangle (1,1);
	\shade[left color=white,right color=gray] (0,-1) rectangle (3.036,0);
	
	\node[no-border] (sequential) at (0,0) {\footnotesize Sequential processors};
	\node[no-border] (multi-core) [right = of sequential] {\footnotesize Multi-core processors};
	\node[no-border] (multi-many-core) [right = of multi-core] {\footnotesize Multi-core and accelerators};
	%Position of sequences
	
	%2000-2005
	%cloud computing
	\node[mh-cell] at (0,0) {};
	%grid computing
	\node[ml-cell] at (0,1) {\footnotesize \cite{lee2003}};
	%single-node
	\node[ml-cell] at (0,2) {\footnotesize \cite{zomaya2001,monsifrot2002,Stephenson2003,zhang2004,Cavazos2004,stephenson2005,cooper2005acme,zhang2005,thomas2005}};
	%multi-node
	\node[ml-cell] at (0,3) {\footnotesize \cite{thomas2005,corbalan2005,page2005,page2005framework}};
	
	%2006-2011
	%cloud computing
	\node[mh-cell] at (1,0) {};
	%grid computing
	\node[ml-cell] at (1,1) {\footnotesize \cite{carretero2007genetic,park2010}};
	%single-node
	\node[ml-cell] at (1,2) {\footnotesize \cite{agakov2006,gordon2006,diamos2008,fursin2008,tournavitis2009,beach2009,chen2009,ansel2009petabricks,luk2009,wang2009,hoffmannESMA2010,hoffmannSEEC2010,eastep2010,pekhimenko2011efficient,grewe2011,castro2011,grewe2011workload,ravi2011,benkner2011peppher,danylenko2011comparing,eastep2011,fursin2011,tiwari2011}};
	%multi-node
	\node[ml-cell] at (1,3) {\footnotesize \cite{cavazos2007rapidly,fursin2008,tiwari2009,sivanandam2009,fursin2011,tiwari2011}};
	
	%2012-2017
	%cloud computing
	\node[mh-cell] at (2,0) {\footnotesize \cite{mantripragada2014,Mastelic2015Cloud,grzonka2017}};
	%grid computing
	\node[ml-cell] at (2,1) {\footnotesize \cite{grzonka2014,gaussier2015}};
	%single-node
	\node[ml-cell] at (2,2) {\footnotesize \cite{kessler2012programmability,li2012adaptive,castro2012,kessler2012optimized,binotto2013,emani2013,albayrak2013,liu2013,wang2013,fonseca2013,ogilvie2015,memeti2016sa,memeti2016saml,memeti2016ml,memeti2016bd,iakymchuk2016particle}};
	%multi-node
	\node[ml-cell] at (2,3) {\footnotesize \cite{rossbach2013,mantripragada2014,li2014,gaussier2015,iakymchuk2016particle,silvano2016}};
	
	\draw (-0.1,1) -- (0,1);
	\draw (-0.1,2) -- (0,2);
	\draw (-0.1,3) -- (0,3);
	\draw (-0.1,4) -- (0,4);
	
	\draw (0,4) -- (0,4.5);
	\draw (1,4) -- (1,4.5);
	\draw (2,4) -- (2,4.5);
	\draw (0,-1) -- (3.036,-1);
	\draw (0,-1) -- (0,0);
	\draw (3.036,-1) -- (3.036,0);
	
	% drawing arrows in the header
	\draw (0.97,-0.8) -- (1.03,-0.5);
	\draw (0.97,-0.2) -- (1.03,-0.5);
	
	\draw (1.94,-0.8) -- (2,-0.5);
	\draw (1.94,-0.2) -- (2,-0.5);
	
	\end{tikzpicture}
	\caption{Classification of state-of-the-art work based on the architecture (multi-node, single-node, grid, and cloud computing systems, as described in Section \ref{sec:parallel-systems}). Please note that a single paper may target more than one architecture (for instance, \cite{mantripragada2014,gaussier2015}).}
	\label{fig:classification-architecture}
\end{figure}

% ------------------------------------------------------------------------------------------------- %
\textbf{Architecture:} Figure \ref{fig:classification-architecture} shows a classification of the reviewed papers based on the target architecture, including multi-node, single-node, grid, and cloud parallel computing systems. The horizontal axis on the top delineates the common types of processors used during the corresponding time period. For instance, from 2000 to 2005 grids and clusters employed single or multiple sequential processors at node level, whereas during the period from 2006 to 2011 nodes employed multi-core processors. Accelerators combined with multi-core processors can be seen during time period 2012-2017. We may observe that most of the work is focused on optimization of resource utilization at the node level (single-node). Optimization of the resources of multi-node computing systems (including clusters) is addressed by several research studies continuously during the considered periods of time. The optimization of grid computing systems using machine learning and meta-heuristic approaches has received less attention, whereas optimization of cloud computing systems has received attention during the period 2012-2017.

% ------------------------------------------------------------------------------------------------- %
\textbf{Software optimization approach:} In Table \ref{tab:software-optimization-classification} we classify the selected publications that use intelligent techniques (such as, machine learning and meta-heuristics) for software optimization at compile-time and run-time. We may observe that machine learning is used more often for software optimization during compile-time and run-time compared to meta-heuristics.

\newcolumntype{P}[1]{>{\centering\arraybackslash}m{#1}}
\begin{table}[]
	\centering
	\caption{Classification of state-of-the-art work based on the intelligent technique (machine learning or meta-heuristics) used during compile-time and/or run-time of software optimization}
	\label{tab:software-optimization-classification}
	\def\arraystretch{1.5}
	\begin{tabular}{l|P{3.2cm}|P{3.3cm}|P{3.9cm}|}
		\cline{1-4}
		\multicolumn{1}{r|}{Machine Learning}
		& \cite{lee2003,monsifrot2002,Cavazos2004,zhang2004,corbalan2005,stephenson2005,thomas2005,zhang2005}
		&
		\cite{agakov2006,cavazos2007rapidly,diamos2008,fursin2008,beach2009,chen2009,tournavitis2009,luk2009,ansel2009petabricks,wang2009,park2010,hoffmannESMA2010,hoffmannSEEC2010,eastep2010,fursin2011,pekhimenko2011efficient,grewe2011,ravi2011,castro2011,grewe2011workload,benkner2011peppher,danylenko2011comparing,eastep2011}
		& \cite{castro2012,kessler2012programmability,li2012adaptive,kessler2012optimized,fonseca2013,liu2013,wang2013,rossbach2013,emani2013,binotto2013,fonseca2013,mantripragada2014,grzonka2014,gaussier2015,ogilvie2015,Mastelic2015Cloud,memeti2016ml,memeti2016bd,memeti2016saml,silvano2016,grzonka2017}
		\\ \hline
		\multicolumn{1}{r|}{Meta-heuristics}
		& \cite{zomaya2001,ahmad2001,zomaya2001introduction,page2005,page2005framework,Stephenson2003,cooper2005acme}
		& \cite{gordon2006,carretero2007genetic,sivanandam2009,tiwari2009,tiwari2011}
		& \cite{albayrak2013,li2014,grzonka2014,memeti2016sa,memeti2016bd,memeti2016saml,grzonka2017}
		\\ \hline
		& 2000-2005 & 2006-2011 & 2012-2017 \\
	\end{tabular}
\end{table}

% ------------------------------------------------------------------------------------------------- %
\textbf{Life-cycle activity:} A classification of the reviewed papers based on the software life-cycle activities (including, code optimization, code generation, scheduling, and adaptation) is depicted in Table \ref{tab:classification-software-lifecycle}. We may observe that the scheduling life-cycle activity has received the most attention, especially during 2012-2017 period. The use of machine learning and meta-heuristics for code optimization during compile-time has been addressed by many researchers, especially during the period between 2006 and 2011. Similar trend can be observed for research studies that focus on using intelligent approaches to optimize code generation. Optimization of software through adaptation is addressed during the year of 2006-2011.

\begin{table}[]
	\centering
	\caption{Classification of state-of-the-art work based on the software life-cycle activities (code optimization, code generation, scheduling, and adaptation). Please note that a single paper may contribute to more than one software life-cycle activities (for instance, \cite{luk2009,beach2009}).}
	\label{tab:classification-software-lifecycle}
	\def\arraystretch{1.5}
	\begin{tabular}{l|P{3.2cm}|P{3.3cm}|P{3.9cm}|}
		\cline{1-4}
		\multicolumn{1}{r|}{Code Optimization}
		& \cite{monsifrot2002,Stephenson2003,Cavazos2004,stephenson2005,cooper2005acme}
		& \cite{agakov2006,gordon2006,cavazos2007rapidly,fursin2008,tournavitis2009,tiwari2009,fursin2011,tiwari2011}
		& \cite{liu2013,wang2013}
		\\ \hline
		\multicolumn{1}{r|}{Code Generation}
		&
		& \cite{beach2009,chen2009,ansel2009petabricks,luk2009,tournavitis2009,pekhimenko2011efficient}
		& \cite{fonseca2013,rossbach2013}
		\\ \hline
		\multicolumn{1}{r|}{Scheduling}
		& \cite{zomaya2001,ahmad2001,zomaya2001introduction,lee2003,zhang2004,corbalan2005,zhang2005,page2005,page2005framework}
		& \cite{diamos2008,wang2009,sivanandam2009,beach2009,park2010,grewe2011,castro2011,grewe2011workload,ravi2011,benkner2011peppher,danylenko2011comparing}
		& \cite{kessler2012optimized,li2012adaptive,kessler2012programmability,castro2012,emani2013,binotto2013,albayrak2013,mantripragada2014,grzonka2014,li2014,gaussier2015,Mastelic2015Cloud,ogilvie2015,memeti2016bd,memeti2016ml,memeti2016sa,memeti2016saml,silvano2016,fonseca2013,grzonka2017}
		\\ \hline
		\multicolumn{1}{r|}{Adaptation}
		& \cite{thomas2005}
		& \cite{luk2009,hoffmannESMA2010,hoffmannSEEC2010,eastep2010,eastep2011}
		&
		\\ \hline
		& 2000-2005 & 2006-2011 & 2012-2017 \\
	\end{tabular}
\end{table}

% ================================================================================================== %
\section{Compile-Time}
\label{sec:compile-time}

Compiling \citep{Aho:2006} is the process of transforming source code from one form into another. Traditionally, compiler engineers exploited the underlying architecture by manually implementing several code transformation techniques. Furthermore, decisions that determine whether to apply a specific optimization or not were hard-coded manually. At each major revision or implementation of new instruction set architecture, the set of such hard-coded compiler heuristics must be re-engineered (a time-consuming process). In the modern era, the architectures are continuously evolving trying to bring higher performance while keeping shorter time to market, therefore developers do not prefer to do the re-engineering, which requires significant time investment.

Modern parallel computing architectures are complex due to higher core counts, different multi-threading, memory hierarchy, computation capabilities, and processor architecture. This disparity of architecture increases the number of available compiler optimization flags and makes compilers unable to efficiently utilize the available resources. Tuning these parameters manually is not just unfeasible, but also introduces scalability and portability issues. Machine learning and meta-heuristics promise to address compiler problems, such as, selecting compiler optimization flags or heuristic-guided compiler optimizations.

In what follows, we discuss the existing state-of-the-art approaches that use machine learning and meta-heuristics for software optimization for code optimization and code generation. Thereafter, we discuss the limitations and identify possible future research directions.

% ------------------------------------------------------------------------------------------------- %
\subsection{Code Optimization}
\label{sec:compile-time-soa-code-optimization}

Code optimization will not change the program behavior but will optimize the code to reach optimization goals (reducing the execution time, energy consumption, or required resources).

Compiler optimization techniques include loop unrolling, splitting and collapsing, instruction scheduling, software pipelining, auto-vectorization, hyper-block formation, register allocation, and data pre-fetching \citep{Stephenson2003}. Different device-specific code optimization techniques may behave differently in various architectures. Furthermore, choosing more than one optimization technique does not necessarily result in better performance, sometimes combination of different techniques may have negative impact on the final output. Hence, manually writing hard-code heuristics is impractical, and techniques that intelligently select the compiler transformations that result in higher application benefits in a given context are required.

Within the scope of this survey, scientific publications that use machine learning for code optimization at compile time include \cite{monsifrot2002,stephenson2005,Cavazos2004,fursin2008,fursin2011,liu2013,wang2013,tournavitis2009,agakov2006}, whereas scientific publications that use meta-heuristics for code optimization include \cite{Stephenson2003,cooper2005acme,tiwari2009,tiwari2011}. Table \ref{tab:code-optimization-characteristics} lists the characteristics of the selected primary studies that address code optimization at compile time. Such characteristics include: the algorithm used for optimization, the optimization objectives, the considered features that describe the application being optimized, and type of optimization (on-line or off-line). We may observe that besides the approach proposed by \citet{tiwari2011}, the rest of them focus on off-line optimization approaches and they are based on historical data (knowledge) that is gathered from previous runs.

\noindent
\fbox{\parbox{\textwidth}{
		\textbf{RQ1: Software optimization goals for compile-time code optimization:}
		\begin{itemize}
			\item loop unrolling; instruction scheduling; partitioning of irregular and stream applications; determining the best compilation parameters; determining whether parallelism is beneficial; tuning compiler heuristics;
		\end{itemize}
}}

As we mentioned earlier, different optimizations can be performed during compilation. We may see that some researchers focus on using intelligent techniques to identify loops that would potentially execute more efficiently when unrolled \cite{monsifrot2002}, or selecting the loop unroll factor that yields the best performance \cite{stephenson2005}. Instruction scheduling \cite{Cavazos2004}, partitioning strategy for irregular \cite{liu2013} and streaming \cite{wang2013} applications, determining the list of compiler optimizations that results in the best performance \cite{fursin2011,cooper2005acme,tiwari2011} are also addressed by the selected scientific publications. Furthermore, \citet{tournavitis2009} use SVMs to determine whether parallelization of the code would be beneficial, and which scheduling policy to select for the parallelized code.

\noindent
\fbox{\parbox{\textwidth}{
		\textbf{RQ2: Software optimization algorithms used for compile-time code optimization:}
		\begin{itemize}
			\item \textit{machine learning} - nearest neighbor classifier; support vector machines; decision trees; ruled set induction; predictive search distribution;
			\item \textit{meta-heuristics} - genetic algorithms; hill climbing; greedy algorithm; parallel rank order;
		\end{itemize}
		
}}

With regards to the machine learning algorithms used for code optimization, Nearest Neighbor (NN) classifier \cite{stephenson2005,liu2013,wang2013,agakov2006}, Support Vector Machine (SVM) \cite{stephenson2005,tournavitis2009}, and Decision Tree (DT) \cite{monsifrot2002} are the most popular. Other algorithms, such as Ruled Set Induction (RSI) \cite{Cavazos2004}, and Predictive Search Distribution (PSD) \cite{fursin2008,fursin2011} are also used for code optimization during compilation. Whereas, approaches that are based on search-based algorithms use Genetic Algorithm (GA), Hill Climbing (HC), Greedy Algorithm (GrA), and Parallel Rank Ordering (PRO) for code optimization during compile-time \cite{cooper2005acme,tiwari2009,tiwari2011}.

\noindent
\fbox{\parbox{\textwidth}{
		\textbf{RQ3: Considered features during compile-time code optimization:}
		
		\begin{itemize}
			\item \textit{loop characteristics} - number of memory accesses, arithmetic operations, statements, loop iterations, floating point operations, operands;
			\item \textit{code-block characteristics} - number of instructions, branches, calls, stores, returns, instructions;
			\item \textit{program features} - type of nested loop; loop bound; loop stride; nest depth;
			\item \textit{static program features} - number of basic blocks in a method, CFG edges, operations, load/store operations; data dependency; loop and branch probability;
			\item \textit{dynamic program features} - number of data accesses, instructions, branches;
			\item \textit{architectural parameters} - cache capacity; register capacity;
			\item \textit{application specific parameters; hyper-block formation features; register allocation features; data pre-fetching features};
		\end{itemize}
}}

To achieve the aforementioned objectives, a representative set of program features is extracted through static code analysis, which are considered to be the most informative with regards to the program behavior. The selection of such features is closely related to the optimization goals. For example, to identify loops that benefit from unrolling, \citet{monsifrot2002} use loop characteristics such as, number of memory accesses, arithmetic operations, code statements, control statements, and loop iterations. Such loop characteristics are also used to determine the loop unroll factor \cite{stephenson2005}. Characteristics related to a specific code block (such as number of instructions, branches, calls, stores) are used when deciding whether applications benefit from instruction scheduling \cite{Cavazos2004}. Determining the partitioning strategy of irregular applications is based on static program features related to basic block, loop characteristics, and the data dependency \cite{liu2013}. Features such as pipeline depth, load/store operations per instruction, number of computations, and computation-communication ratio are used when determining partitioning strategy of streaming applications \cite{wang2013}. \citet{tiwari2011} consider architectural specifications such as cache and register capacity, in addition to the application specific parameters, such as tile size in a matrix multiplication algorithm.

\newcommand{\ignore}[1]{}

\begin{table}[]
	\centering
	\scriptsize
	\caption{Characteristics of the approaches that use machine learning or meta-heuristics for code optimization. Please note that, because of space limitation, we do not list all of the considered optimization features.}
	\label{tab:code-optimization-characteristics}
	\begin{tabular}{@{}p{0.5cm} p{1.1cm} p{3.5cm} p{6.4cm} p{1.5cm} @{}}
		\toprule
		Paper
		& Algorithm
		& Objectives
		& Features
		& On/Off-line
		\\ \toprule
		\iffalse Paper \fi 			  \cite{monsifrot2002}
		\iffalse Algorithm	\fi		& DT
		\iffalse Objectives \fi 	& identify loops to unroll
		\iffalse Features \fi		& loop characteristics (\# memory accesses; \# arithmetic operations; \# statements; \# control statements; \# iterations)
		\iffalse On/Off-line \fi 	& off-line (sup.)
		\\ \midrule
		
		\iffalse Paper \fi 			  \cite{stephenson2005}
		\iffalse Algorithm	\fi		& NN, SVM
		\iffalse Objectives \fi 	& select the most beneficial loop unroll factor
		\iffalse Features \fi		& loop characteristics (\# floating point operations; \# operands; \# memory operations; critical path length; \# iterations) 
		\iffalse On/Off-line \fi 	& off-line (sup.)
		\\ \midrule
		
		\iffalse Paper \fi 			  \cite{Cavazos2004}
		\iffalse Algorithm	\fi		& RSI
		\iffalse Objectives \fi 	& determine whether to apply instructions scheduling
		\iffalse Features \fi		& code-block characteristics (\# instructions; \# branches; \# calls; \# stores; \# returns; int/float/sys\_func\_unit instructions)
		\iffalse On/Off-line \fi 	& off-line (sup.)
		\\ \midrule
		
		\iffalse Paper \fi 			  \cite{fursin2008,fursin2011}
		\iffalse Algorithm	\fi		& PSD
		\iffalse Objectives \fi 	& determine the most effective compiler optimizations
		\iffalse Features \fi		& static program features (\# basic blocks in a method; \# normal/critical/abnormal CFG edges;  \# int/float operations)
		\iffalse On/Off-line \fi	& off-line (sup.)
		\\ \midrule
		
		\iffalse Paper \fi 			  \cite{liu2013}
		\iffalse Algorithm	\fi		& kNN
		\iffalse objectives \fi 	& determine the best partitioning strategy of irregular applications
		\iffalse Features \fi		& static program features (\# basic blocks; \# instructions; loop probability; branch probability; data dependency)
		\iffalse On/Off-line \fi 	& off-line (sup.)
		\\ \midrule
		
		\iffalse Paper \fi 			  \cite{wang2013}
		\iffalse Algorithm	\fi		& NN
		\iffalse Objectives \fi 	& determine the best partitioning strategy of streaming app.
		\iffalse Features \fi		& program features (pipeline depth; split-join width; pipeline/split-join work; \# computations; \# load/store ops)
		\iffalse On/Off-line \fi 	& off-line (sup.)
		\\ \midrule
		
		\iffalse Paper \fi 			  \cite{tournavitis2009}
		\iffalse Algorithm	\fi		& SVM
		\iffalse Objectives \fi 	& determine whether parallelism is beneficial; select the best scheduling policy
		\iffalse Features \fi		& static program features (\# instructions; \# load/store; \# branches; \# iterations); dynamic program features (\# data accesses; \# instructions; \# branches)
		\iffalse On/Off-line \fi 	& off-line (sup.)
		\\ \midrule
		
		\iffalse Paper \fi 			  \cite{agakov2006}
		\iffalse Algorithm	\fi		& IIDM; MM; NN;
		\iffalse Objectives \fi 	& reduce the number of required program evaluations in iterative compiler optimization; analyze program similarities
		\iffalse Features \fi		& program features (type of nested loop; loop bound; loop stride; \# iterations; nest depth; \# array references; \# instructions; \# load/store/compare/branch/divide/call/generic/array/memory copy/other instructions; int/float variables)
		\iffalse On/Off-line \fi	& off-line (sup.)
		\\ \midrule
		
		\iffalse Paper \fi 			  \cite{Stephenson2003}
		\iffalse Algorithm	\fi		& GP
		\iffalse Objectives \fi 	& tuning compiler heuristics
		\iffalse Features \fi		& hyper-block formation features; register allocation features; data pre-fetching features.
		\iffalse On/Off-line \fi	& off-line (unsupervised)
		\\	 \midrule
		
		\iffalse Paper \fi 			  \cite{cooper2005acme}
		\iffalse Algorithm	\fi		& GrA; GA; HC; RP;
		\iffalse Objectives \fi 	& tuning the compilation process through adaptive compilation
		\iffalse Features \fi		& /
		\iffalse On/Off-line \fi 	& off-line
		\\ \midrule
		
		\iffalse Paper \fi 			  \cite{tiwari2009,tiwari2011}
		\iffalse Algorithm	\fi		& PRO
		\iffalse Objectives \fi 	& tune generated code; determine the best compilation parameters
		\iffalse Features \fi		& architectural parameters (cache capacity; register capacity); application specific parameters
		\iffalse On/Off-line \fi 	& on-line
		\\ \bottomrule
	\end{tabular}
\end{table}

% ------------------------------------------------------------------------------------------------- %
\subsection{Code Generation}
\label{sec:compile-time-soa-code-generation}

The process of transforming code from one representation into another one is called code generation. We call ``machine code generation'' the code transformation from the high level to low level representation (that is ready for execution), whereas ``source code generation'' indicates in this paper the source-to-source code transformation.

In the context of parallel computing, a source-to-source compiler is an automatic parallelization compiler that can automatically annotate a sequential code with parallel code annotations (such as, OpenMP pragma directives or MPI code statements). Source-to-source compilers may alleviate the portability issue, by enabling to automatically translate the code into an equivalent representation of the code that is ready to be compiled and executed on target architectures.

In this section, we focus on source code generation techniques that can:
\begin{itemize}
	\item generate device-specific code from other code representations,
	\item generate multiple implementations of the same code, or
	\item automatically generate parallel code from sequential code.
\end{itemize}

During the process of porting applications, programmers are faced with the following problems: (1) demand of device-specific knowledge and API; (2) difficulties to predict whether the  application will have performance benefits before it is ported; (3) there exist a large number of programming languages and models that are device (types and manufacturer) specific.

To address such issues, researchers have proposed different solutions. In Table \ref{tab:code-generation-characteristics}, we list the characteristics of these solutions such as, optimization algorithm, optimization objectives, and considered features during optimization.

\noindent
\fbox{\parbox{\textwidth}{
		\textbf{RQ1: Software optimization goals for compile-time code generation:}
		\begin{itemize}
			\item generating device-specific code; mapping applications to accelerating devices; generating multi-threaded loop versions; source-to-source transformations; determining the list of program method transformations; enabling writing multiple versions of algorithms and algorithmic choices at language level; auto-tuning algorithmic choices and switching between them during program execution; determining optimal work distribution between CPU and GPU.
		\end{itemize}
}}

The optimization objectives are derived from the aforementioned portability challenges. For example, to alleviate the demand for device-specific knowledge, \citet{beach2009} aim to identify candidate kernels that would likely benefit from parallelization, generate device-specific code from high-level code, and map to the accelerating device that yields the best performance. Similarly, \citet{fonseca2013} propose the automatic generation of OpenCL code from Java code. \citet{ansel2009petabricks} propose the PetaBricks framework that enables writing multiple versions of algorithms, which are automatically translated into C++ code. The runtime can switch between the available algorithms during program execution. \citet{luk2009} introduce Qilin that enables source-to-source transformation from C++ to TBB and CUDA. It uses machine learning to find the optimal work distribution between the CPU and GPU on a heterogeneous system.

\noindent
\fbox{\parbox{\textwidth}{
		\textbf{RQ2: Software optimization algorithms used for compile-time code generation:}
		\begin{itemize}
			\item \textit{machine learning} - decision trees; near neighbors; linear regression;
		\end{itemize}
		
}}

Decision Trees (DT) \cite{beach2009, fonseca2013}, k-Nearest Neighbor (kNN) \cite{chen2009}, Cost Sensitive Decision Table (CSDT), Naive Bayes (NB), Support Vector Machine (SVM), Multi-layer Perceptron (MPL) \cite{fonseca2013}, Linear Regression (LR) \cite{luk2009, fonseca2013}, and Logistic Regression (LRPR) \cite{pekhimenko2011efficient} machine learning algorithms are used during the code-generation.

\noindent
\fbox{\parbox{\textwidth}{
		\textbf{RQ3: Considered features during compile-time code generation:}
		
		\begin{itemize}
			\item \textit{loop characteristics} - data precision; amount of computation performed; memory access characteristics; loop type; loop statement
			\item \textit{general program features} - number of instructions; load/store operations; floating point operations
			\item \textit{static code features} - loop nest depth; number of arrays; outer/inner access/write; basic operations;
			\item \textit{dynamic features}	- data set size; data-to; data-from;
			\item \textit{runtime algorithm parameters; hardware configuration parameters}
			
		\end{itemize}
}}

\citet{beach2009} considered static loop characteristics to achieve their objectives, whereas \citet{chen2009} use both static and dynamic program features to generate the multi-threaded versions of a selected loop, and then select the most suitable loop version at run-time. Combination of static code features (extracted at compile time), and dynamic features (extracted at run-time) are also used to determine the most suitable processing device for a specific application \cite{fonseca2013}. To determine the best workload distribution of a parallel application, \citet{luk2009} consider algorithm parameters and hardware configuration parameters. \citet{pekhimenko2011efficient} consider general and loop-based features to determine the list of program method transformation during code generation that would reduce the compilation time.

\begin{table}[]
	\centering
	\scriptsize
	\caption{Characteristics of the approaches that use machine learning or meta-heuristics for code generation.}
	\label{tab:code-generation-characteristics}
	\begin{tabular}{@{}p{0.3cm} p{1cm} p{5cm} p{5cm} p{1.5cm} @{}}
		\toprule
		Paper
		& Algorithm
		& Objectives
		& Features
		& On/Off-line
		\\ \toprule
		\iffalse Paper \fi 			  \cite{beach2009}
		\iffalse Algorithm	\fi		& DT
		\iffalse Objectives \fi 	& generate device-specific code from high-level code; map applications to accelerating devices.
		\iffalse Features \fi		& loop (kernel) characteristics (data precision, amount of computation performed and memory access characteristics)
		\iffalse On/Off-line \fi 	& off-line (sup.)
		\\ \midrule
		
		\iffalse Paper \fi 			  \cite{chen2009}
		\iffalse Algorithm	\fi		& kNN
		\iffalse Objectives \fi 	& generate multi-threaded loop versions; select the most suitable one at run-time
		\iffalse Features \fi		& static code features (loop nest depth, \# arrays used); dynamic features (data set size)
		\iffalse On/Off-line \fi 	& off-line (sup.)
		\\ \midrule
		
		\iffalse Paper \fi 			  \cite{fonseca2013}
		\iffalse Algorithm	\fi		& NB, SVM, MPL, CSDT, LR
		\iffalse Objectives \fi 	& source-to-source transformation of data-parallel applications; predict the efficiency and select the suitable device.
		\iffalse Features \fi		& static program features (outer/inner access/write; basic operations; ...); dynamic program features (data-to; data-from; ...)
		\iffalse On/Off-line \fi 	& off-line (sup.)
		\\ \midrule
		
		\iffalse Paper \fi 			  \cite{ansel2009petabricks}
		\iffalse Algorithm	\fi		& /
		\iffalse Objectives \fi 	& enable writing multiple versions of algorithms and algorithmic choices at the language level; auto-tuning of the specified algorithmic choices; switch between the available algorithms during program execution
		\iffalse Features \fi		& /
		\iffalse On/Off-line \fi 	& off-line
		\\ \midrule
		
		\iffalse Paper \fi 			  \cite{luk2009}
		\iffalse Algorithm	\fi		& LR
		\iffalse Objectives \fi 	& determine the optimal work distribution between the CPU and GPU
		\iffalse Features \fi		& runtime algorithm parameters (input size) and hardware configuration parameters
		\iffalse On/Off-line \fi 	& on-line
		\\ \midrule
		
		\iffalse Paper \fi 			  \cite{rossbach2013}
		\iffalse Algorithm	\fi		& /
		\iffalse Objectives \fi 	& distribute data-parallel portions of a program across heterogeneous computing resources;
		\iffalse Features \fi		& /
		\iffalse On/Off-line \fi 	& /
		\\ \midrule
		
		\iffalse Paper \fi 			  \cite{pekhimenko2011efficient}
		\iffalse Algorithm	\fi		& LRPR
		\iffalse Objectives \fi 	& determine the list of program method transformations that result in lower compilation time
		\iffalse Features \fi		& general program features (\# instructions; \# load/store operations; \# float operations); loop-based features (\# loops types; \# loop statements)
		\iffalse On/Off-line \fi 	& off-line (sup.)
		\\ \midrule
	\end{tabular}
\end{table}

% ------------------------------------------------------------------------------------------------- %
\subsection{Observations, Challenges, and Future Directions}
\label{sec:compile-time-crd}

In this section, we first discuss the advantages of meta-heuristics and machine learning methods for software optimization at compile-time, followed by a discussion about their limitations. Thereafter, we discuss the future directions.

In table \ref{tab:achievements-compile-time}, we list each of the machine learning and meta-heuristic methods used for compile-time software optimization. For each of the used methods, we provide the advantages, such as performance improvement, speedup, and prediction accuracy.

\begin{table}[]
	\centering
	\scriptsize
	\caption{Advantages of meta-heuristics and machine learning methods for compile-time software optimization}
	\label{tab:achievements-compile-time}
	\begin{tabular}{@{}p{0.2cm} p{3cm} p{10.5cm} @{}}
		\toprule
		& Method & Advantages \\ \midrule
		\multirow{9}{*}{\rotatebox[origin=c]{90}{\parbox[c]{5cm}{\centering Machine Learning}}}
		
		& Decision Trees & \cite{monsifrot2002} reports up to 3\% performance improvement for loop unrolling; \cite{beach2009} reports performance achievement within 15\% of the performance achieved manually-ported code.\\ \cmidrule{2-3}
		
		& Support Vector Machines  & \cite{stephenson2005} report that SVM and NN can predict the optimal unroll factor for a given loop 65\% of the time or the near optimal one 79\% of the time. \cite{tournavitis2009} uses SVM to decide whether to parallelize loop candidates and achieves 96\% of the performance of hand-tuned code. \cite{fonseca2013} report 92\% of prediction accuracy when using SVM to decide whether kernels should be executed on GPU or CPU. \\ \cmidrule{2-3}
		
		& (k) Nearest Neighbor 		& \cite{liu2013} report 5.41\% performance improvement compared to hard-coded compiler optimizations.  \cite{wang2013} use NN to predict the partitioning structure of applications. The authors achieve up to 1.9 $\times$ speedup compared to the default partitioning strategy, which is 60\% of the ideal one.  \cite{chen2009} report that 87\% of the highest performance improvement can be achieved using NN. \\ \cmidrule{2-3}
		
		& Ruled Set Induction 		& \cite{Cavazos2004} use RSI to determine whether or not to apply instruction scheduling on a code block and reported achievement of 90\% performance improvement compared to schedule always method. \\ \cmidrule{2-3}
		
		& Regression Based Algorithms & \cite{pekhimenko2011efficient} use regression techniques to determine the optimal heuristic parameters and report two fold speedup of the compilation process while maintaining the same code quality. \\ \cmidrule{2-3}
		
		& Decision Tables 			& In \cite{fonseca2013}, the up to 92 \% prediction accuracy of DT, NB, and MLP to decide whether kernels should be executed on the GPU or CPU is significant to achieve 65x speedup over sequential Java programs. \\ \cmidrule{2-3}
		
		& Predictive Search Distribution & \cite{fursin2008,fursin2011} use PSD to select the best compiler optimizations and report 11\% performance improvement. \\ \midrule

		\multirow{4}{*}{\rotatebox[origin=c]{90}{\parbox[c]{1.8cm}{\centering Meta-heuristics}}}
		
		& Greedy Algorithm and Hill Climbing & \cite{cooper2005acme} use meta-heuristics to find the optimal compilation parameters while reducing the number of evaluations during search space exploration from 10000 to a single one using profiling data and estimated virtual execution. \\ \cmidrule{2-3}
		
		& Genetic Algorithm & \cite{Stephenson2003} obtain speedup of 23\% for hyper-block formation.\\ \cmidrule{2-3}
		
		& Parallel Rank Ordering & \cite{tiwari2009,tiwari2011} use PRO for automatic tuning of compilation process and report 46\% performance improvement compared to the original code. \\ \bottomrule
		
	\end{tabular}
\end{table}

\begin{table}[]
	\centering
	\scriptsize
	\caption{Limitations of the existing studies that use machine learning and meta-heuristics for compile-time software optimization}
	\label{tab:limitations-compile-time}
	\begin{tabular}{@{}p{1.2cm} p{3cm} p{9.3cm}@{}}
		\toprule
		Studies & Focus & Limitations\\ \midrule
		
		\cite{monsifrot2002,stephenson2005,Cavazos2004,Stephenson2003} & Single aspects of code optimization & Control single/few and simple optimization (such as: loop unrolling; instruction scheduling; hyper-block formation). Considering multiple and more complex compiler optimizations is more challenging. \\ \midrule
		
		\cite{fursin2008,fursin2011,tiwari2009,tiwari2011,pekhimenko2011efficient} & Determining the most effective compiler optimization & Training is based in random data sampling and it requires large number of samples, which may reduce its effectiveness. \\ \midrule
		
		\cite{liu2013,wang2013, luk2009} & Determining the best partitioning strategy & Assumes that for any two functions with similar features, the same partitioning strategy can be used. \\ \midrule
		
		\cite{tournavitis2009} & Determining loops that benefit from parallelization and their best scheduling policy & Targets OpenMP loop constructs only. Uses profiling to detect loop candidates, which may significantly increase the compilation time. \\ \midrule
		
		\cite{cooper2005acme,agakov2006} & Adaptive tuning of the compilation process & Profiling data needs to be collected to perform the virtual executions. Takes too long to find the optimal transformations. Efficient for simple models, but results in lower prediction accuracies for more complex problems. \\ \midrule
		
		\cite{beach2009,chen2009,fonseca2013} & Device-specific code generation; mapping applications to accelerators & The prediction model requires significant training data for accurate mapping decisions. Lack of training data may result in performance degradation.\\ \midrule
		
		\cite{chen2009} & Generating multi-threaded versions of the loop & The size of the executable file may dramatically increase for applications with large parallel code, and hardware architectures that consist of multiple multi-core processing units. \\ \midrule
		
		\cite{fonseca2013,rossbach2013,ansel2009petabricks} & Source-to-source transformation & Limited to map-reduce operations. The automatic code generation is limited to specific features of Java code. Not all Java code can be translated to OpenCL \\ \midrule
		
		\cite{ansel2009petabricks} & Auto-tuning; run-time library & PetaBricks requires that developers write their application using non widely known programming languages. The performance of PetaBricks is closely dependent on the architecture where the auto-tuning was performed. \\ \bottomrule
		
	\end{tabular}
\end{table}

While most of the approaches discussed in this review present significant performance improvement, which is important towards having intelligent compilers that require less engineering effort to provide satisfactory code execution performance, indications that there is still room for improvement can be observed in \citet{Stephenson2003} and \citet{wang2013}.

Limitations of the compile-time software optimization approaches that use machine learning or meta-heuristics are listed in Table \ref{tab:limitations-compile-time}, which include: (1) limitation to a specific programming language or model \citep{tournavitis2009}, (2) forcing developers to use extra annotations on their code \citep{luk2009}, or use not widely known parallel programming languages \citep{ansel2009petabricks}, (3) focusing on single or simpler aspects of optimizations techniques (ex: loop unrolling, unrolling factor, instruction scheduling) \citep{monsifrot2002,stephenson2005,Cavazos2004}, whereas more complex compiler optimizations (that are compute-intensive) are not addressed sufficiently.

Furthermore, optimizations based on features derived from static code analysis provide poor global characterization of the dynamic behavior of the applications, whereas using dynamic features requires application profiling, which adds additional execution overhead to the program under study. This additional time can be considered negligible for applications that are executed multiple times after the optimization, however it represents overhead for single-run applications. Approaches that generate many multi-threaded versions of the code  \cite{chen2009} might end up with dramatic code increases that make difficult the applicability to embedded parallel computing systems with limited resources. Adaptive compilation techniques \citep{cooper2005acme} add non-negligible compilation overhead.

Future research should address the identified shortcomings in this systematic review by providing intelligent compiler solutions for general-purpose languages (such as, C/C++) and compilers (for instance, GNU Compiler Collection) that are widely used and supported by the community. Many compiler optimization issues are complex and require human resources that are usually not available within a single research group or project.

% ================================================================================================= %
\section{Run-Time}
\label{sec:run-time}

The run-time program life-cycle is the time during which the program is running (that is, being executed) and it is also known as execution-time. Software systems that enable running programs to interact with the execution environment are known as run-time systems. The run-time environment contains environment information, such as, the available resources, existing workload, and scheduling policy. A running program can access the execution environment information via the run-time system.

In the past, the choice of architecture and the algorithms was considered during the design and implementation phase of software life-cycle. Nowadays, there are various multi- and many-core processing devices, with different performance and energy consumption characteristics. Furthermore, there is no single algorithm implementation that can exploit the full processing potential of these diverse processing elements. Often it is not possible to know if an application performs better on device X or Y before the execution. The performance of a program is determined by the properties of the execution context (program input, type of available processing elements, current system utilization...) that is known at run-time. Some programs perform better on device X when the input size is large enough, but worse for smaller input sizes. Hence, decisions whether a program should be run on X or Y, or which algorithm to use are postponed to run-time.

In this study, we focus on optimization methods used in different run-time systems that use machine learning or meta-heuristics to optimize the program execution. Such run-time systems may be responsible for partitioning programs into tasks and scheduling these tasks to different processing devices, selecting the most suitable device(s) for a specific task, selecting the most suitable algorithm or the size of the input workload, selecting the number of processing elements or clock frequency, and many more different system run-time configuration parameters to achieve the specified goals including the performance, energy efficiency, and fault tolerance. Specifically, we focus on two major run-time activities: \emph{scheduling} and \emph{adaptation}.

In what follows, we discuss the related state-of-the-art run-time optimization approaches for scheduling and adaptation. Thereafter, we summarize the limitations of the current approaches and discuss possible future research directions.

% ------------------------------------------------------------------------------------------------- %
\subsection{Scheduling}
\label{sec:run-time-soa-scheduling}

According to the Cambridge Dictionary \footnote{Cambridge Dictionary, \url{http://dictionary.cambridge.org/dictionary/english/scheduling}}, scheduling is \enquote{the job or activity of planning the times at which particular tasks will be done or events will happen}. In context of this paper, we use the term scheduling to indicate mapping the tasks onto the processing elements and determining the order of task execution to minimize the overall execution time.

Scheduling may strongly influence the performance of parallel computing systems. Improper scheduling can lead to load imbalance and consequently to sub-optimal performance. Researchers have proposed different approaches that use meta-heuristics or machine learning to find the best scheduling within a reasonable time.

Based on whether the scheduling algorithms can modify the scheduling policy during program execution, generally scheduling algorithms are classified in static and dynamic.

% ------------------------------------------------------------------------------------------------- %
\subsubsection{Static Scheduling}

Static scheduling techniques retain an unchanged policy until the end of program execution. Static approaches assume that the number of tasks is fixed, known before execution starts, and that accurate information of their running times is known. Static approaches usually use analytical models to estimate the computation and communication cost, where the work distribution is performed based on these estimations.
The program execution time is essential for job scheduling. However, accurately predicting/estimating the program execution time is difficult to achieve in shared environments where system resources can dynamically change over time. Inaccurate predictions may lead to performance degradation \cite{chirkin2017}.

\begin{table}[]
	\centering
	\scriptsize
	\caption{Characteristics of the approaches that use machine learning or meta-heuristics for static scheduling.}
	\label{tab:static-scheduling-characteristics}
	\begin{tabular}{@{}p{0.8cm} p{1cm} p{4cm} p{5.2cm} p{1.8cm} @{}}
		\toprule
		Paper
		& Algorithm
		& Objectives
		& Features
		& On/Off-line
		\\ \toprule
		
		\iffalse Paper \fi 			  \cite{wang2009}
		\iffalse Algorithm	\fi		& ANN; SVM
		\iffalse Objectives \fi 	& mapping computations to multi-core CPUs; determine the optimal thread number;
		\iffalse Features \fi		& code features (\# static instructions; \# load/store operations; \# branches); data and dynamic features (L1 data cache miss rate; branch miss rate)
		\iffalse On/Off-line \fi 	& off-line (sup.)
		\\ \midrule
		
		\iffalse Paper \fi 			  \cite{grewe2011}
		\iffalse Algorithm	\fi		& SVM
		\iffalse Objectives \fi 	& mapping computations to the suitable processing device
		\iffalse Features \fi		& static code features (\# int/float/math operations; barriers; memory accesses; \% local/coalesced memory accesses; compute-memory ratio)
		\iffalse On/Off-line \fi 	& off-line (sup.)
		\\ \midrule
		
		\iffalse Paper \fi 			  \cite{castro2011}
		\iffalse Algorithm	\fi		& ID3 DT
		\iffalse Objectives \fi 	& mapping threads to specific cores; reduce memory latency and contention
		\iffalse Features \fi		& program features (transaction time ratio; transaction abort ratio; conflict detection policy; conflict resolution policy; cache misses)
		\iffalse On/Off-line \fi 	& off-line (sup.)
		\\ \midrule
		
		\iffalse Paper \fi 			  \cite{ogilvie2015}
		\iffalse Algorithm	\fi		& L; MP; IB1; IBk; KStar ...
		\iffalse Objectives \fi 	& reducing the training data; select the most informative training data; mapping application to processors;
		\iffalse Features \fi		& /
		\iffalse On/Off-line \fi 	& off-line (sup.)
		\\ \midrule
		
		\iffalse Paper \fi 			  \cite{memeti2016ml}
		\iffalse Algorithm	\fi		& BDTR
		\iffalse Objectives \fi 	& determine workload distribution of data-parallel applications on heterogeneous systems
		\iffalse Features \fi		& hardware configuration (\# threads; \# cores; \# threads/core; thread affinity); application parameters (input size)
		\iffalse On/Off-line \fi 	& off-line (sup.)
		\\ \midrule
		
		\iffalse Paper \fi 			  \cite{memeti2016sa}
		\iffalse Algorithm	\fi		& SA
		\iffalse Objectives \fi 	& determine near-optimal system configuration parameters of heterogeneous systems
		\iffalse Features \fi		& system configuration parameters (\#threads/thread\_affinity/ workload\_fraction on host/device);
		\iffalse On/Off-line \fi 	& off-line (sup.)
		\\ \midrule
		
		\iffalse Paper \fi 			  \cite{memeti2016bd,memeti2016saml}
		\iffalse Algorithm	\fi		& BDTR; SA
		\iffalse Objectives \fi 	& determine near-optimal system configuration on heterogeneous systems
		\iffalse Features \fi		& available resources; scheduling policy; and the workload fraction;
		\iffalse On/Off-line \fi 	& off-line (sup.)
		\\ \midrule
		
		\iffalse Paper \fi 			  \cite{zomaya2001,ahmad2001,zomaya2001introduction,carretero2007genetic}
		\iffalse Algorithm	\fi		& GA
		\iffalse Objectives \fi 	& task scheduling
		\iffalse Features \fi		& /
		\iffalse On/Off-line \fi 	& off-line (sup.)
		
		\\ \bottomrule
	\end{tabular}
\end{table}

Table \ref{tab:static-scheduling-characteristics} lists the characteristics (such as optimization algorithm, objective, and features) of scientific publications that use machine learning and/or meta-heuristics for static scheduling.

\noindent
\fbox{\parbox{\textwidth}{
		\textbf{RQ1: Software optimization goals for run-time static scheduling:}
		\begin{itemize}
			\item mapping program parallelism to multi-core architectures; determining the optimal number of threads; mapping applications to the most appropriate processing device; reducing memory latency and contention; mapping threads to specific cores; determining workload distribution on heterogeneous systems; determining near-optimal system configuration parameters;
		\end{itemize}
}}

With regards to static scheduling, the attention of recent research that use machine learning and meta-heuristics is in the following optimization objectives: mapping program parallelism to multi-core architectures \cite{wang2009}, mapping applications to the most appropriate processing device \cite{grewe2011,ogilvie2015}, mapping threads to specific cores \cite{castro2011}, and determining workload distribution on heterogeneous parallel computing systems \cite{memeti2016ml,memeti2016sa,memeti2016bd,memeti2016saml}.

\noindent
\fbox{\parbox{\textwidth}{
		\textbf{RQ2: Software optimization algorithms used for run-time static scheduling:}
		\begin{itemize}
			\item \textit{machine learning} - artificial neural networks; support vector machines; (boosted) decision trees; logistic; multi-layer perceptron; IB1; IBk; KStar; Random Forest; LogitBoost; multiclass classifier; NNge; ADTree; random tree;
			\item \textit{meta-heuristics} - simulated annealing; genetic algorithms;
		\end{itemize}
}}

To achieve the aforementioned optimization objectives, machine learning algorithm such as, Artificial Neural Networks (ANN), Support Vector Machines (SVM), and (Boosted) Decision Trees (BDTR) are used \cite{wang2009,grewe2011,castro2011,memeti2016ml}. An approach that combines a number of machine learning algorithms, including, Logistic (L), Multilayer Perceptron (MP), IB1, IBk, KStar, Random Forest, Logit Boost, Multi-Class-Classifier, Random Committee, NNge, ADTree, and RandomTree, to create an active-learning query-committee with the aim to reduce the required amount if training data is proposed by \citet{ogilvie2015}. A combination of Simulated Annealing (SA) and boosted decision tree regression to determine near optimal system configurations is proposed by \citet{memeti2016saml}. The use of Genetic Algorithms (GA) for task scheduling has been extensively addressed by several researchers \citep{zomaya2001,ahmad2001,zomaya2001introduction,carretero2007genetic}.

\noindent
\fbox{\parbox{\textwidth}{
		\textbf{RQ3: Considered features during run-time static scheduling:}
		\begin{itemize}
			\item \textit{static program features} - number of static instructions; number of load/store operations; number of branches; barriers; memory accesses; compute-memory ratio; transaction time ratio; transaction abort ratio; conflict detection and resolution policy;
			\item \textit{data and dynamic features} - L1 data cache miss rate; branch miss rate;
			\item \textit{hardware characteristics} - number of threads, cores, threads per core; thread affinity;
			\item \textit{system configuration parameters} - input size; workload fraction on host and accelerating devices;
		\end{itemize}
}}

The list of considered system features for optimizing of parallel computing systems is closely related to the optimization objectives, target applications and architecture. For example, \citet{castro2011} consider transaction time and abort ratio, conflict detection and resolution policy to map thread to specific cores and reduce memory latency and contention in software transactional memory applications running on multi-core architectures. Static code features, such as number of instruction, memory operations, math operations, and branches, are considered during the mapping of applications to the most suitable processing devices \cite{wang2009,grewe2011}. While such approaches consider application specific features, researchers have demonstrated positive improvement results in approaches that do not require code analysis. Instead, they rely on features such as the available system resources and program input size during the optimization process (that is determining the workload distribution of data-parallel applications) \cite{memeti2016sa,memeti2016saml,memeti2016ml}.

% ------------------------------------------------------------------------------------------------- %
\subsubsection{Dynamic Scheduling}

Dynamic scheduling algorithms take into account the current system state and modify themselves during run-time to improve the scheduling policy. Dynamic scheduling does not require prior knowledge of all task properties. To overcome the limitations of the static scheduling, various dynamic approaches are proposed, including work stealing, partitioning and assigning tasks on the fly, queuing systems, and task-based approaches. Dynamic scheduling is usually harder to implement; however, the performance gain may be better than static scheduling.

\begin{center}
	\scriptsize
	\begin{longtable}{@{}p{0.5cm} p{1cm} p{4.4cm} p{5.4cm} p{1.5cm} @{}}
		\caption{Characteristics of the approaches that use machine learning or meta-heuristics for dynamic scheduling.}
		\label{tab:dynamic-scheduling-characteristics} \\
		\toprule
		Paper
		& Algorithm
		& Objectives
		& Features
		& On/Off-line
		\\ \toprule
		
		\endfirsthead
		
		Paper
		& Algorithm
		& Objectives
		& Features
		& On/Off-line
		\\ \toprule
		\endhead
		
		\iffalse Paper \fi 			  \cite{emani2013}
		\iffalse Algorithm	\fi		& ANN
		\iffalse Objectives \fi 	& determine the best number of threads
		\iffalse Features \fi		& static features (\# load/store ops; \# instructions; \# branches); dynamic features (\# processors; \# workload threads; run queue length; ldavg-1; ldavg-5)
		\iffalse On/Off-line \fi 	& off-line (sup.)
		\\ \midrule
		
		\iffalse Paper \fi 			  \cite{lee2003}
		\iffalse Algorithm	\fi		& R\&F %regression and filtering
		\iffalse Objectives \fi 	& determine the application execution time in shared environments
		\iffalse Features \fi		& program input parameters; \# processors; resource status;
		\iffalse On/Off-line \fi 	& off-line (sup.)
		\\ \midrule
		
		\iffalse Paper \fi 			  \cite{park2010}
		\iffalse Algorithm	\fi		& SVM
		\iffalse Objectives \fi 	& mapping tasks to processing devices
		\iffalse Features \fi		& \# tasks in the queue; the ready times of the machines; computing capabilities of each machine.
		\iffalse On/Off-line \fi 	& off-line (sup.)
		\\ \midrule
		
		\iffalse Paper \fi 			  \cite{mantripragada2014}
		\iffalse Algorithm	\fi		& GrA
		\iffalse Objectives \fi 	& evenly partitioning tasks between high performance clusters and the cloud
		\iffalse Features \fi		& estimated execution time determined by monitoring the actual exec. time of data or tasks chunks
		\iffalse On/Off-line \fi 	& on-line
		\\ \midrule
		
		\iffalse Paper \fi 			  \cite{Mastelic2015Cloud}
		\iffalse Algorithm	\fi		& /
		\iffalse Objectives \fi 	& predicting resource allocation for business processes in the Cloud
		\iffalse Features \fi		& runtime metrics of a process and its behavior
		\iffalse On/Off-line \fi 	& off-line
		\\ \midrule
		
		\iffalse Paper \fi 			  \cite{castro2012}
		\iffalse Algorithm	\fi		& ID3 DT
		\iffalse Objectives \fi 	& predicting a thread mapping strategy for STM applications
		\iffalse Features \fi		& Transactional Time/Abort Ratio; Conflict Detection/Resolution Policy; Last-Level Cache Miss
		\iffalse On/Off-line \fi 	& off-line (sup.)
		\\ \midrule
		
		\iffalse Paper \fi 			  \cite{grzonka2014}
		\iffalse Algorithm	\fi		& ANN
		\iffalse Objectives \fi 	& improve the effectiveness of grid scheduler decisions % regarding resource allocations with the security condition considered as the most important factor
		\iffalse Features \fi		& characteristics of the tasks and machines
		\iffalse On/Off-line \fi 	& off-line
		\\ \midrule
		
		\iffalse Paper \fi 			  \cite{grzonka2017}
		\iffalse Algorithm	\fi		& ANN; GA
		\iffalse Objectives \fi 	& improve the makespan
		\iffalse Features \fi		& security demands, workload of task, and the output size 
		\iffalse On/Off-line \fi 	& off-line \& on-line
		\\ \midrule
		
		\iffalse Paper \fi 			  \cite{gaussier2015}
		\iffalse Algorithm	\fi		& LR
		\iffalse Objectives \fi 	& improving the scheduling algorithms using machine learning techniques
		\iffalse Features \fi		& job arrival time; required resources; \# running jobs; occupied resources;
		\iffalse On/Off-line \fi 	& on-line
		\\ \midrule
		
		\iffalse Paper \fi 			  \cite{binotto2013}
		\iffalse Algorithm	\fi		& /
		\iffalse Objectives \fi 	& optimize the task scheduling on heterogeneous platforms
		\iffalse Features \fi		& input data; data transfers; task performance; platform features;
		\iffalse On/Off-line \fi 	& off-line \& on-line
		\\ \midrule
		
		\iffalse Paper \fi 			  \cite{grewe2011workload}
		\iffalse Algorithm	\fi		& ANN
		\iffalse Objectives \fi 	& predict the optimal number of threads
		\iffalse Features \fi		& program features and workload features
		\iffalse On/Off-line \fi 	& off-line
		\\ \midrule
		
		\iffalse Paper \fi 			  \cite{zhang2004,zhang2005}
		\iffalse Algorithm	\fi		& Adaptive LR % Based Adaptive; Loop-based; Hardware-Counter Directed
		\iffalse Objectives \fi 	& determine the number of threads and scheduling policy for each parallel region
		\iffalse Features \fi		& inter-thread data locality, instruction mix and load imbalance
		\iffalse On/Off-line \fi 	& /
		\\ \midrule
		
		\iffalse Paper \fi 			  \cite{page2005,page2005framework}
		\iffalse Algorithm	\fi		& GA
		\iffalse Objectives \fi 	& minimize the make-span; dynamic task scheduling in heterogeneous systems
		\iffalse Features \fi		& task properties(arrival time; dependency); system properties (network; processors)
		\iffalse On/Off-line \fi 	& on-line
		\\ \midrule
		
		\iffalse Paper \fi 			  \cite{albayrak2013}
		\iffalse Algorithm	\fi		& Adaptive GrA
		\iffalse Objectives \fi 	& mapping of computation kernels on heterogeneous GPUs accelerated systems.
		\iffalse Features \fi		& profiling information (execution time; data-transfer time); hardware characteristics
		\iffalse On/Off-line \fi 	& off-line
		\\ \midrule
		
		\iffalse Paper \fi 			  \cite{li2014}
		\iffalse Algorithm	\fi		& HC
		\iffalse Objectives \fi 	& selecting optimal per task system configuration for MapReduce applications
		\iffalse Features \fi		& map-reduce parameters (\# mappers; \# reducers; slow start; io.sort.mb; \# virtual cores)
		\iffalse On/Off-line \fi 	& on-line
		\\ \midrule
		
		\iffalse Paper \fi 			  \cite{sivanandam2009}
		\iffalse Algorithm	\fi		& PSO; SA
		\iffalse Objectives \fi 	& dynamic scheduling of heterogeneous tasks on heterogeneous processors; load balancing;
		\iffalse Features \fi		& task properties (execution time; communication cost; fitness function); hardware properties (\# processors)
		\iffalse On/Off-line \fi 	& /
		\\ \midrule
		
		\iffalse Paper \fi 			  \cite{zomaya2001}
		\iffalse Algorithm	\fi		& GA
		\iffalse Objectives \fi 	& dynamic load-balancing where optimal task scheduling can evolve at run-time
		\iffalse Features \fi		& /
		\iffalse On/Off-line \fi 	& on-line
		\\ \midrule
		
		\iffalse Paper \fi 			  \cite{ravi2011}
		\iffalse Algorithm	\fi		& /
		\iffalse Objectives \fi 	& mapping tasks to heterogeneous architectures; % predict the optimal size chunk of applications with different problem sizes and system configurations
		\iffalse Features \fi		& architectural trade-offs; computation patterns; application characteristics;
		\iffalse On/Off-line \fi 	& on-line
		\\ \midrule
		
		\iffalse Paper \fi 			  \cite{diamos2008}
		\iffalse Algorithm	\fi		& PR
		\iffalse Objectives \fi 	& dynamic scheduling and performance optimization for heterogeneous systems
		\iffalse Features \fi		& kernel execution time; machine parameters; input size; input distribution var.; instrumentation data;
		\iffalse On/Off-line \fi 	& off-line (supervised)
		\\ \midrule
		
		\iffalse Paper \fi 			  \cite{benkner2011peppher,kessler2012programmability}
		\iffalse Algorithm	\fi		& LR; QR
		\iffalse Objectives \fi 	& prediction of performance aspects (e.g. execution time, power consumption) of implementation variants;
		\iffalse Features \fi		& system information (resource availability and requirements; estimated performance of implementation variants; input availability)
		\iffalse On/Off-line \fi 	& off-line (supervised)
		\\ \midrule
		
		\iffalse Paper \fi 			  \cite{li2012adaptive}
		\iffalse Algorithm	\fi		& DT
		\iffalse Objectives \fi 	& reducing the number of training data required to build prediction models
		\iffalse Features \fi		& input parameters (e.g. size); system available resources (e.g. \# cores; \# accelerators);
		\iffalse On/Off-line \fi 	& off-line (supervised)
		\\ \midrule
		
		\iffalse Paper \fi 			  \cite{kessler2012optimized,danylenko2011comparing}
		\iffalse Algorithm	\fi		& DT; DD; NB; SVM
		\iffalse Objectives \fi 	& use meta-data from performance aware components to predict the expected execution time; select the best implementation variant and the scheduling policy;
		\iffalse Features \fi		& input parameters (e.g. size); system available resources; meta-data
		\iffalse On/Off-line \fi 	& off-line (supervised)
		\\ \bottomrule
	\end{longtable}
\end{center}

Table \ref{tab:dynamic-scheduling-characteristics} lists the characteristics (such as optimization algorithm, objective, and features) of scientific publications that use machine learning and/or meta-heuristics for dynamic scheduling.

\noindent
\fbox{\parbox{\textwidth}{
		\textbf{RQ1: Software optimization goals for run-time dynamic scheduling:}
		\begin{itemize}
			\item mapping tasks to processing devices; partitioning tasks between performance clusters and the cloud; determining resource allocation; predicting thread mapping strategy; predicting the optimal number of threads; improving scheduling algorithms; determining scheduling policy; minimizing the make-span; mapping computation kernels to heterogeneous GPU accelerated systems; determining optimal system configuration; load balancing; determining performance aspects, such as execution time ad power consumption; selecting the best algorithm implementation variant; reducing the number of training data required to build prediction models;
		\end{itemize}
}}

With regards to the optimization objectives, considered scientific publications aim at: (1) determining the optimal number of threads for a given application \cite{emani2013,grewe2011workload,zhang2004,zhang2005}; (2) determining the application execution time \cite{lee2003,benkner2011peppher,kessler2012programmability,kessler2012optimized,danylenko2011comparing}; (3) mapping tasks to processing devices \cite{park2010,castro2011,albayrak2013,ravi2011}; (4) partitioning tasks between high performance clusters \cite{mantripragada2014}; (5) predicting resource allocation in the cloud \cite{Mastelic2015Cloud}; (6) improving scheduling algorithms \cite{grzonka2014,gaussier2015}; (7) minimizing the make-span \cite{binotto2013,page2005,page2005framework,sivanandam2009,zomaya2001,diamos2008,grzonka2017}; (8) selecting near optimal system configurations \cite{li2014}; and (9) reducing the number of training examples required to build prediction models \cite{li2012adaptive}.

\noindent
\fbox{\parbox{\textwidth}{
		\textbf{RQ2: Software optimization algorithms used for run-time dynamic scheduling:}
		\begin{itemize}
			\item \textit{machine learning} - artificial neural networks; regression and filtering techniques; support vector machines; (boosted) decision trees; logistic; multi-layer perceptron; IB1; IBk; KStar; Random Forest; LogitBoost; multiclass classifier; NNge; ADTree; random tree; dispatch tables; Naive Bayesian classifier; decision diagrams;
			\item \textit{meta-heuristics} - (adaptive) greedy algorithm; simulated annealing; genetic algorithms; hill climbing; particle swarm optimization;
		\end{itemize}
}}

Artificial neural network (ANN) \cite{emani2013,grzonka2014,grewe2011workload,grzonka2017}, regression (LR, QR, PR) \cite{lee2003,gaussier2015,diamos2008,benkner2011peppher,kessler2012programmability}, support vector machines (SVM) \cite{park2010,kessler2012optimized,danylenko2011comparing}, and decision trees (DT) \cite{castro2012,li2012adaptive,kessler2012optimized,danylenko2011comparing} are the most popular machine learning algorithms used for optimization in the scientific publications considered in this study. Whereas, genetic algorithms (GA) \cite{page2005,page2005framework,zomaya2001,grzonka2017}, greedy-based algorithms (GrA) \cite{mantripragada2014,albayrak2013}, hill-climbing (HC) \cite{li2014}, particle swarm optimization (PSO) \cite{sivanandam2009}, and simulated annealing (SA) \cite{sivanandam2009} are used as heuristic based optimization approaches for dynamic scheduling.

\noindent
\fbox{\parbox{\textwidth}{
		\textbf{RQ3: Considered features during run-time dynamic scheduling:}
		\begin{itemize}
			\item \textit{static and dynamic features} - number of load/store operations; number of instructions and branches; number of processors and workload threads; run-queue length;
			\item \textit{task related features} - number of tasks in the queue; machine ready time; estimated task execution time; task performance; arrival time; dependency;
			\item \textit{runtime information} - metrics of a process and its behavior; last level cache misses; job arrival time; running jobs; data transfers; inter-thread data locality; instruction mix and load imbalance; execution time; data-transfer time; fitness function;
			\item \textit{application specific and workload characteristics} - transactional time and abort ratio; conflict detection and resolution policy; required resources; input data; number of mappers and reducers in map-reduce applications;
			\item \textit{hardware characteristics} - machine computing capability; occupied resources; platform features; network properties; processor properties;
		\end{itemize}
}}

Approaches such as \cite{mantripragada2014,castro2012,Mastelic2015Cloud,li2014,zhang2004,zhang2005} focus on features collected dynamically during program execution, such as, estimated execution time determined through analysis of profiling data, information related to tasks (arrival time, number of currently running tasks). Whereas other approaches combine static features collected at compile-time with dynamic ones collected at run-time \cite{emani2013,park2010,grzonka2014,page2005,page2005framework}, program input parameters, and hardware related information \cite{binotto2013,lee2003,grewe2011workload,ravi2011,diamos2008}. Similar to the static scheduling techniques, the selection of such features is closely related to the optimization objectives. For example, \citet{zhang2004,zhang2005} consider the inter-thread data locality when tuning OpenMP applications for hyper-threaded SMPs; \citet{page2005,page2005framework} consider task properties, such as, task arrival time and task dependency, when scheduling dynamically tasks in heterogeneous distributed systems. Features such as security demands, workload of tasks, and the output size are considered to train the ANN for optimization of scheduling process and maximization of resource usage in the cloud \cite{grzonka2017}.

% ------------------------------------------------------------------------------------------------- %
\subsection{Adaptation}
\label{sec:run-time-soa-adaptive}

According to the Cambridge Dictionary \footnote{Cambridge Dictionary, \url{http://dictionary.cambridge.org/dictionary/english/adaptation}}, adaptation is \enquote{the process of changing to suit different conditions.} In this paper, we use the term adaptation to refer to the property of systems that are capable of evaluating and changing their behavior to achieve specified goals with respect to performance, energy efficiency, or fault tolerance. In dynamic environments, modern parallel computing systems may change their behavior by: (1) changing the number of used processing elements to optimize system resource allocation; (2) changing the algorithm or implementation variant that yields to better results with respect to the specified goals; (3) reducing the quality (accuracy) of the output to meet the performance goals; or (4) changing the clock frequency to reduce energy consumption.

\begin{table}[]
	\centering
	\scriptsize
	\caption{Characteristics of the optimization approaches based on adaptation techniques.}
	\label{tab:adaptation-characteristics}
	\begin{tabular}{@{}p{0.4cm} p{1.9cm} p{3.2cm} p{4.2cm} p{3.3cm} @{}}
		\toprule
		Paper
		& Method
		& Adaptation Objectives
		& Monitored Parameters
		& Tuned Parameters
		
		\\ \toprule
		\iffalse Paper \fi 			  		\cite{thomas2005}
		\iffalse Method	\fi					& Custom adaptation loop; DT
		\iffalse Objectives \fi 			& select the most suitable algorithm implementation
		\iffalse Monitoring parameters \fi	& architecture and system information (available memory, cache size, \# processors);  performance characteristics
		\iffalse Tuned parameters \fi 		& algorithm implementation
		\\ \midrule
		
		\iffalse Paper \fi 			  		\cite{hoffmannESMA2010,hoffmannSEEC2010}
		\iffalse Method	\fi					& ODA
		\iffalse Objectives \fi 			& apply user defined actions to change the program behavior 
		\iffalse Monitoring parameters \fi	& performance information retrieved using application heartbeats \cite{hoffmannESMA2010}
		\iffalse Tuned parameters \fi 		& user defined actions (such as: adjusting the clock speed) 
		\\ \midrule
		
		\iffalse Paper \fi 			  		\cite{eastep2010}
		\iffalse Method	\fi					& Lock Acquisition Scheduling; RL
		\iffalse Objectives \fi 			& adapt the lock's internal implementation 
		\iffalse Monitoring parameters \fi	& reward signal (heart rate) retrieved using application heartbeats.
		\iffalse Tuned parameters \fi 		& change the lock scheduling policy
		\\ \midrule
		
		\iffalse Paper \fi 			  		\cite{eastep2011}
		\iffalse Method	\fi					& RL 
		\iffalse Objectives \fi 			& determine the ideal  data structure knob settings
		\iffalse Monitoring parameters \fi	& reward signal (throughput heart rate); support for external perf. monitors
		\iffalse Tuned parameters \fi 		& adjusting the \textit{scancount} value and performance-critical knob of Flat Combining algorithm.
		\\ \midrule
		
		\iffalse Paper \fi 			  		\cite{luk2009}
		\iffalse Method	\fi					& LR
		\iffalse Objectives \fi 			& adaptive mapping of computations to PE 
		\iffalse Monitoring parameters \fi	& execution-time of parts of the program
		\iffalse Tuned parameters \fi 		& choosing the mapping scheme (static or adaptive)
		\\ \midrule
		
		\iffalse Paper \fi 			  		\cite{silvano2016}
		\iffalse Method	\fi					& DSL
		\iffalse Objectives \fi 			& adapt applications to meet user defined goals 
		\iffalse Monitoring parameters \fi	& contextual information, requirements, resources availability 
		\iffalse Tuned parameters \fi 		& user defined actions (altering resource alloc. and task mapping)
		
		\\ \bottomrule
	\end{tabular}
\end{table}

The studied literature in this paper provide examples that adaptation (also referred to as self-adaptation) proved to be an effective approach to deal with the complexity, variability, and dynamism of modern parallel computing systems. Table \ref{tab:adaptation-characteristics} lists the characteristics (such as, adaptation method and objectives, monitored and tuned parameters) of the scientific publications that use adaptation for software optimization of parallel computing systems.

\noindent
\fbox{\parbox{\textwidth}{
		\textbf{RQ1: Software optimization goals for run-time adaptation:}
		\begin{itemize}
			\item selecting the most suitable algorithm implementation; applying user defined actions to change the program behavior; adapting lock's internal implementation mechanisms; determining the ideal data structure knob settings; adaptive mapping of computations to processing elements; adapting applications to meet the user defined goals;
		\end{itemize}
}}

With regards to the adaptation objectives, \citet{thomas2005} use a custom adaptation loop to adaptively select the most suitable algorithm implementation for a given input data set and system configuration. \citet{hoffmannESMA2010,hoffmannSEEC2010} use an observe-decide-act (ODA) feedback loop to adaptively apply user defined actions to change the program behavior in favor of achieving some user-defined goals, such as energy efficiency and throughput. Adaptation methods are used in the smart-locks library \cite{eastep2010}, which can change its behavior at run-time to achieve certain goals. Similarly, in \cite{eastep2011} adaptation methods are used for optimizing data structure knobs. Adaptive mapping of computations to the processing units is proposed by \citet{luk2009}. The Antarex \citep{silvano2016} project aims at providing means for application tuning and adaptation for energy efficient heterogeneous high-performance computing systems, by providing a domain specific language that allows specifying adaptation goals at compile-time.

\noindent
\fbox{\parbox{\textwidth}{
		\textbf{RQ2: Software optimization algorithms used for run-time adaptation:}
		\begin{itemize}
			\item \textit{machine learning} - decision trees; reinforcement learning; linear regression;
			\item \textit{other} - custom adaptation loop; observe-act-decide loops; lock acquisition scheduling;
		\end{itemize}
}}

During the process of adaptation, all of the approaches proposed in the considered scientific publications, have at least three components of an adaptation loop, including monitoring, deciding, and acting. For example, \citet{thomas2005} monitor architecture and environment parameters, then uses a decision tree to analyze such information, and perform the required changes (in this case selecting an algorithm implementation). Similarly, \citet{hoffmannSEEC2010} use the so called observe-decide-act (ODA) feedback loop to monitor performance related information (retrieved using the application heartbeats \cite{hoffmannESMA2010}) and use the heart-rate to take some user defined actions, such as adjusting the clock speed, allocating cores, or change the algorithm. Reinforcement learning (RL), an on-line machine learning algorithm, is used to help with the adaptation decisions in both smart-locks \cite{eastep2010} and smart data-structures \cite{eastep2011}, whereas linear regression (LR) is used by \citet{luk2009} for choosing the mapping scheme of computations to processing elements.

\noindent
\fbox{\parbox{\textwidth}{
		\textbf{RQ3: Considered features during run-time dynamic scheduling:}
		\begin{itemize}
			\item \textit{hardware characteristics} - available memory; cache size; number of processors; resource availability
			\item \textit{performance characteristics} - heartbeat reward signal; throughput; external performance monitors; execution time;
			\item \textit{contextual information; requirements; }
		\end{itemize}
}}

In Table \ref{tab:adaptation-characteristics} we list two types of parameters, the monitored parameters, used to evaluate whether adaptation goals have been met, and tuned parameters, which are basically defined actions that will change the program behavior until the desired goals are achieved. For monitoring, architecture and environment variables (such as, available memory, cache size, number of processors), and performance characteristics are considered by \citet{thomas2005}. Performance related information retrieved from the heartbeats monitor are used as monitoring parameters in the following scientific articles \cite{hoffmannSEEC2010,eastep2010,eastep2011}. \citet{luk2009} rely on the execution time of parts of the program, whereas the Antarex framework uses contextual information, requirements, and resource availability for monitoring the program behavior. As tuning parameters, the following are considered, selecting the algorithm implementation \cite{thomas2005,hoffmannSEEC2010}, adjusting the clock speed, core allocation, select algorithm \cite{hoffmannSEEC2010}, change lock scheduling policy \cite{eastep2010}, adjust the scancount \cite{eastep2011}, change mapping scheme \cite{luk2009}, and altering resource allocation and task mapping \cite{silvano2016}.

% ------------------------------------------------------------------------------------------------- %
\subsection{Observations, Challenges and Research Directions}
\label{sec:run-time-crd}

In this section, we first discuss the advantages of meta-heuristics and machine learning methods for software optimization at run-time, followed by a discussion about their limitations. Thereafter, we discuss the future directions.

In table \ref{tab:achievements-run-time}, we list each of the machine learning and meta-heuristic methods used for run-time software optimization. For each of the used methods, we provide the advantages, including performance improvement, speedup, or prediction accuracy.

\begin{table}[]
	\centering
	\scriptsize
	\caption{Advantages of meta-heuristics and machine learning methods for run-time software optimization}
	\label{tab:achievements-run-time}
	\begin{tabular}{@{}p{0.2cm} p{3cm} p{10.5cm} @{}}
		\toprule
		& Method & Advantages \\ \midrule
		\multirow{9}{*}{\rotatebox[origin=c]{90}{\parbox[c]{8.5cm}{\centering Machine Learning}}}
		
		& Artificial Neural Network	& \cite{emani2013} report speedup of up to $3.2\times$ compared to OpenMP default scheme, and $2.3\times$ compared to Hill Climbing on-line adaptation technique.  \cite{grzonka2014} show that the ANN can be used to reduce the time required to find the best possible solutions by approximately 30-40\%. \cite{grewe2011workload} show that their neural network is aware of existing workload and can reduce the slowdown to existing workload from 4.5\% to 0.5\% at a cost of reducing the speedup from $1.66\times$ to $1.59\times$.  \\ \cmidrule{2-3}
		
		& Support Vector Machines & \cite{grewe2011} report performance achievement of 80.6\% compared to the optimal one. \cite{wang2009} use ANN and SVM to determine the best number of threads and show performance achievements of up to 96\% compared to the optimal performance. \cite{kessler2012optimized} show that the SVMs can be used to select the best optimization variant with 0\% inaccuracy, however the decision overhead is high.  \\ \cmidrule{2-3}
		
		& Decision Trees	& \cite{castro2011} show performance improvement of up to 18.46\% compared to the worst case scenario. \cite{memeti2016ml} can determine a near-optimal workload distribution on heterogeneous system, which results in performance improvement of up to 35.6x compared to sequential version. \cite{thomas2005} show that a performance accuracy between 86-100\% is capable to dynamically optimize the execution time by choosing the most suitable algorithm in a given context. \\ \cmidrule{2-3} %\cite{li2012adaptive} show that the prediction accuracy depends on the depth of the dispatch trees, where increase of the depth results in more accurate predictions, however it increases the training time.
		
		& Regression	& \cite{gaussier2015} can predict the execution time, which help to achieve up to 28\% makespan reduction. \cite{zhang2005} show performance improvement up to 27\% when using regression techniques to predict the optimal number of threads and scheduling policy. \cite{luk2009} use regression techniques to map computations to processing units, which result in performance improvement up to ~40\% compared to mapping always to CPU, 25\% compared to GPU-always, and within 94\% of the near optimal mapping.  \\ \cmidrule{2-3}
		
		& Reinforcement Learning	& \cite{eastep2010} reported up to 1.2x speedup compared to other approaches for lock acquisition scheduling. \cite{eastep2011} show the ability to adapt scancount to changing application needs, which result in up to 1.5x speedup compared to state-of-the-art approaches. \\ \midrule

		\multirow{4}{*}{\rotatebox[origin=c]{90}{\parbox[c]{4cm}{\centering Meta-heuristics}}}
		
		& Simulated Annealing	& \cite{memeti2016saml} use simulated annealing to optimize the workload distribution on heterogeneous systems. By evaluating only about 5\% of all possible configurations it can achieve average speedup of 1.6x and 2x compared with the host-only and device-only execution.  \\ \cmidrule{2-3}
		
		& Genetic Algorithms	& \cite{zomaya2001} show that GA performs better than First Fit for dynamic scheduling using various number of tasks and available processing elements. \cite{page2005,page2005framework} show that their evolutionary based scheduler outperforms other schedulers.  \\ \cmidrule{2-3}
		
		& Greedy Algorithm	& \cite{mantripragada2014} predicts the application execution time, and allows to dynamically shift part of the workload from the cluster to be computed in the cloud, in order to meet the deadline. \cite{albayrak2013} show that nine out of ten times the mapping algorithm based on GrA performs better than single-device mapping.  \\ \cmidrule{2-3}
		
		& Hill Climbing	& \cite{li2014} shows performance improvement of up to 30\% compared to the default configurations used by YARN. \\ \cmidrule{2-3}
		
		& Particle Swarm Optimization & \cite{sivanandam2009} uses PSO and SA for task scheduling. The hybridization of these algorithms outperforms other algorithms, including GA. \\ \bottomrule
	\end{tabular}
\end{table}

\begin{table}[]
	\centering
	\scriptsize
	\caption{Limitations of the existing studies that use machine learning and meta-heuristics for run-time software optimization}
	\label{tab:limitations-run-time}
	\begin{tabular}{@{}p{1.2cm} p{3cm} p{9.3cm}@{}}
		\toprule
		Studies & Focus & Limitations\\ \midrule
		
		\cite{wang2009,grewe2011} & Mapping computations to the most suitable processing units  & The mapping process is dependent on the hardware architecture, which means that a mapping scheme that fits well an architecture may not yield the desired performance on another architecture. It requires to re-learn the prediction model for each new application and architecture configuration. \\ \midrule
		
		\cite{wang2009,grewe2011workload,zhang2004,zhang2005} & Determining the optimal number of threads  & Require off-line training. \cite{grewe2011workload} adapts to the workload only when the application starts its execution, but it does not adapt throughout its execution. \cite{zhang2004} focus on applications that consist of single-loops. \\ \midrule
		
		\cite{castro2011}	& Determining thread mapping strategy for TM applications	& Since it uses static features, it can not change the mapping strategy when the parallelism degree changes at runtime. \\ \midrule
		
		\cite{li2014}	& Determining near-optimal system configuration parameters	& Requires extensive profiling data collection and analysis, which may introduce significant run-time overhead. \\ \midrule
		
		\cite{page2005,page2005framework,diamos2008}	& Dynamic task scheduling	& The proposed approaches do not consider dependencies between tasks. \\ \midrule
		
		\cite{ravi2011}	& Dividing tasks into chunks and scheduling in task-farm way	& Task-farm or master-slave like scheduling techniques requires no profiling, however they introduce communication overheads. \\ \midrule
		
		\cite{zomaya2001,zomaya2001introduction,ahmad2001}	& Task scheduling	& Assume that communication time is known prior execution, processing units have equal computing power and are always ready to perform tasks, and scheduling can be determined off-line and it can not be changed at run-time. \\ \midrule
		
		\cite{ogilvie2015} & Reducing the number of required training data	& This solution requires additional programming investment to build the prediction committee. \\ \midrule
		
		\cite{albayrak2013} & Mapping of computation to heterogeneous systems  & Collecting profiling information for each kernel on every device represent huge overhead, especially on heterogeneous systems that may comprise larger number of non-identical CPUs and/or GPUs.  \\ \midrule
		
		\cite{thomas2005}	& Adaptive algorithm selection & Considers hardware characteristics, but not the program input characteristics. Some algorithms perform better for smaller input sizes, whereas others perform better with larger ones.\\ \midrule
		
		\cite{hoffmannESMA2010,hoffmannSEEC2010,eastep2010,eastep2011,silvano2016,luk2009}	& Self-adaptation	&	Require adding additional information to the code. Require running the application for a certain amount of time (often with non-optimal parameters) until the framework takes a more optimal decision.\\ \bottomrule
		
	\end{tabular}
\end{table}

At run-time, many execution environment parameters influence the performance behavior of the program. Exploring this huge parameter space is time consuming and impractical for programs with long execution times and large demand for system resources. Different computing capabilities and energy efficiency of processing elements of heterogeneous parallel computing systems make the scheduling a difficult challenge. Table \ref{tab:limitations-run-time} lists the limitations of the run-time software optimization approaches for parallel computing systems considered in this paper.

We may observe that some of the existing scheduling techniques often assume that the program is executed on a dedicated system and all system resources are available for use. The approach proposed by \citet{grewe2011workload} propose a co-scheduling technique, which considers that the resources are shared with other applications. However, the adaptation occurs only when the application is executed, but not during program execution. We believe that better results could be achieved if they consider to adapt to changes while the application is being executed. Another issue is that commonly used scheduling techniques ignore slow processing elements due to their low performance capabilities. Mapping computations always to processing units that offer higher performance capability is not optimal, because slower processing elements may never get work to perform. Furthermore, most of the reviewed approaches target specific features of the code only (for example, loops), or are limited to specific programming models and applications (data-bound or compute-bound). Many static scheduling approaches require retraining of the prediction model for each new architecture, limiting their general use because training requires a significant amount of data that is not always available. Approaches that reduce the amount of training data require implementation of multiple machine learning algorithms (for instance, \citep{ogilvie2015}). Approaches that use a single execution \citep{li2014,cooper2005acme} by trying various system configurations during the program execution are promising, however the introduced overhead is not negligible. Self-adaptation techniques require the developer to add additional information into the code so that the software would be able to monitor the system and take decisions. Even though such code is not difficult to add for the application programmer, the software development becomes more complex while talking decisions based on these results. Furthermore, such approaches introduce overhead at runtime, because they need to run for a certain amount of time until enough data is collected for the framework to be able to take the most optimal decisions.

Future research should aim at reducing the scheduling and adaption overhead for dynamic approaches. Run-time optimization techniques for heterogeneous systems should be developed that utilize all available computing resources to achieve the optimization goals. There is a need for robust run-time optimization frameworks that are useful for a large spectrum of programs and system architectures. Furthermore, techniques that reduce the amount of data generated from system monitoring are needed in particular for extreme-scale systems.

% ================================================================================================= %
\section{Conclusion}
\label{sec:conclusion}

In this article, we have conducted a systematic literature review that describes approaches that use machine learning and meta-heuristics for software optimization of parallel computing systems. We have classified approaches based on the software life-cycle activities at compile-time and run-time, including the code optimization and generation, scheduling, and adaptation. We have discussed the shortcomings of existing approaches and provided recommendations for future research directions.

\begin{table}[]
	\centering
	\scriptsize
	\caption{Advantages and limitations of software optimization approaches that use machine learning and meta-heuristics}
	\label{tab:conclusion-overview}
	\begin{tabular}{@{}p{0.2cm} p{6.75cm} p{6.75cm} @{}}
		\toprule
		& Advantages & Limitations \\ \midrule
		\multirow{3}{*}{\rotatebox[origin=c]{90}{\parbox[c]{3cm}{\centering Compile-time}}}
		
		& \begin{itemize}
			\item[+] Source-to-source code generation and optimization may outperform manually written code.
			\item[+] Selecting the best compiler optimizations results in better performance compared to hard-coded compiler optimizations.
			\item[+] In comparison to techniques that assume that each code block benefits from certain optimizations, performance improvement is observed for approaches that intelligently determine for each code block whether such optimizations result in better performance.
		\end{itemize}	
		& \begin{itemize}
			\item Some approaches focus on single and simple compiler optimizations. Considering multiple and more complex compiler optimizations is more challenging.
			\item Training requires large amount of data or extensive profiling of the code. Lack of training data may result in performance degradation.
			\item Some approaches generate multiple versions of the code, and select the most optimal one at runtime. The code-size may dramatically increase.
		\end{itemize}	
		\\ \midrule
		
		\multirow{3}{*}{\rotatebox[origin=c]{90}{\parbox[c]{3cm}{\centering Run-time}}}
		
		& \begin{itemize}
			\item[+] Co-scheduling can reduce the slowdown of other applications.
			\item[+] Selecting the best system configurations or algorithm implementation variant for a given execution context may result in better performance and reduce the time required for application tuning.
			\item[+] Adaptation approaches may change the program behavior during execution to meet certain user-defined goals.
		\end{itemize}	
		& \begin{itemize}
			\item Some of approaches consider only few aspects of the system, such as code features, but not hardware characteristics, system utilization, or task dependencies.
			\item Some approaches assume that applications will be executed in isolation, whereas real-world applications may share the same resources with other applications.
			\item Some approaches ignore slow processing elements, which may result in overall system underutilization. %Mapping computations always to faster processing elements is not optimal, because slower processing elements may never get work to perform.
		\end{itemize}	
		\\ \bottomrule
	\end{tabular}
\end{table}

A high-level overview is provided in Table \ref{tab:conclusion-overview}, which lists the advantages and limitations of the compile-time and run-time software optimization approaches that use machine learning and meta-heuristics.
Our analysis of the reviewed literature suggests that the use of machine learning and meta-heuristic based techniques for software optimization of parallel computing systems is capable of delivering performance comparable to the manual code optimization or task scheduling strategies in specific cases. However, many existing solutions are limited to a specific programming language and model, type of application, or system architecture. There is a need for software optimization frameworks that are applicable to a large spectrum of programs and system architectures. Future efforts should focus on developing solutions for widely used general-purpose languages (such as, C/C++) and compilers that are used and supported by the community.

% ================================================================================================= %

%	\footnotesize
%
%	\bibliographystyle{spbasic}
%	\bibliography{slr,compiler,runtime}
%}

\end{document}